\documentclass[10pt,emulateapj,apj]{emulateapj}

\shorttitle{The Horizon Run}
\shortauthors{Kim, et al.}
\begin{document}
\title{The Horizon Run N-body Simulation: Baryon Acoustic Oscillations and \
Topology of Large Scale Structure of the Universe}
\author{ Juhan Kim\altaffilmark{1}, Changbom Park\altaffilmark{2}, 
J. Richard Gott III\altaffilmark{3}, \& John Dubinski\altaffilmark{4}}
\altaffiltext{1}{Canadian Institute for Theoretical Astrophysics, University of Toronto, 60 St. George Street,
Toronto, ON M5S 3H8, Canada; kjhan@cita.utoronto.ca}
\altaffiltext{2}{Korea Institute for Advanced Study, Dongdaemun-gu, Seoul 130-722, Korea; cbp@kias.re.kr}
\altaffiltext{3}{Department of Astrophysical Sciences, Princeton University, Princeton, NJ08550, USA}
\altaffiltext{4}{Department of Astronomy \& Astrophysics, University of Toronto, 50 St. George St., Toronto, Ontario M5S 3H4, Canada}

\begin{abstract}
In support of the new Sloan III survey, which will measure the baryon oscillation 
scale using the luminous red galaxies (LRGs), we have run the largest N-body simulation to date
using $4120^3 = 69.9$ billion particles, and covering a volume of 
$(6.592~ h^{-1} {\rm Gpc})^3$.
This is over 2000 times the volume of the Millennium Run, 
and corner-to-corner stretches all the way to the horizon of the visible universe.   
LRG galaxies are selected by finding the most massive gravitationally bound, 
cold dark matter subhalos, not subject to tidal disruption, a technique 
that correctly reproduces the 3D topology of the LRG galaxies in the Sloan 
Survey.  
We have measured the covariance function, power spectrum, and the 3D
topology of the LRG galaxy distribution
in our simulation and made 32 mock surveys along the past light 
cone to simulate the Sloan III survey.  
Our large N-body simulation is used
to accurately measure the non-linear systematic effects
such as gravitational evolution, redshift space
distortion, past light cone space gradient, and galaxy biasing, and
to calibrate the baryon oscillation scale and the genus topology.
For example,
we predict from our mock surveys that the baryon acoustic oscillation
peak scale can be measured with
the cosmic variance-dominated uncertainty of about 5\% when the SDSS-III sample
is divided into three equal volume shells, or about 2.6\% when a thicker shell
with $0.4<z<0.6$ is used.  We find that one needs to correct the scale for the
systematic effects amounting up to 5.2\% to use it to constrain the
linear theories.
And the uncertainty in the amplitude of the genus curve is expected to be about 1\%
at 15 $h^{-1}$Mpc scale.
We are making the simulation and mock surveys publicly available.
\end{abstract}

\keywords{cosmological parameters -- cosmology:theory -- large-scale structure of universe -- galaxies: formation -- methods: n-body simulations}

\section{Introduction}

A leading method for characterizing dark energy is measuring the baryon 
oscillation scale using Luminous Red Galaxies (LRGs).  Baryon oscillations 
are the method ``least affected by systematic uncertainties'' according 
to the Dark Energy task Force report (Albrecht et al. 2006).  The baryon 
oscillation scale provides a ``standard ruler'' (tightly constrained by 
the WMAP observations), which has been detected as a bump in 
the covariance function of the LRG galaxies in the Sloan Survey 
(and also in the 2dFGRS survey) (see Percival et al. 2007 for a review). 

This observation has prompted a team to propose the Baryon Oscillation 
Spectroscopic Survey (BOSS) for an extension of the Sloan Survey (Sloan III), 
which will obtain the redshifts of 1,300,000 LRG galaxies
over 10,000 square degrees selected from the Sloan 
digital image survey.  This will make it possible to measure 
the baryon bump in the covariance function of the LRG galaxies to a high 
accuracy as a function of redshift out to a redshift of z = 0.6.  
It is claimed that this allows, in principal, 
measurement of the Hubble parameter to an accuracy of 1.7\% at $z=0.6$, 
and allows us to place narrower constraints on $w(z)$ (the ratio of dark energy 
pressure to energy density as a function of redshift).
Systematic uncertainties are claimed
to be of the order of 0.5\% (Eisenstein et al 2007$a$; Crocce \& 
Scoccimarro 2007).  These systematics are only as well controlled as knowledge 
of all non-linear effects associated with LRG galaxies and thus 
one needs large N-body simulations.  Small N-body simulations--$(256)^3$ particles, 
($512 h^{-1}$Mpc)$^3$ (Seo \& Eisenstein 2005), show that non-linear 
effects do change the covariance function of the LRG galaxies relative 
to that in the initial conditions.  The baryon bump in the covariance 
function is somewhat washed out, but can be recovered 
by reconstruction techniques moving the galaxies backward to
their initial conditions using the Zeldovich approximation 
(Eisenstein et al. 2007$b$).  However, these simulations 
are inadequate because of their small box size which is only a factor 
of 4.7 larger than the baryon oscillation scale which is of order of 
$108 h^{-1}$Mpc.   It is very important to  model the power 
spectrum accurately at large scales and to have a large box size so that 
the statistical errors in the power spectrum are small.  After all, we need 
to measure the acoustic peak scale to an accuracy better than 1\%.  

Concerning the 
Baryon Acoustic Oscillation method applied to dark energy, Crotts et al. 
(2005) note that ``systematic effects will 
inevitably be present in real data; they can only be addressed by means of 
studying mock catalogues constructed from realistic cosmological volume 
simulations'' just as we are setting about to do here.  
According to Crotts et al. (2005) such modeling would typically improve the 
accuracy of the estimated dark energy parameters by 30-50\%.  
In support of this goal we have previously completed a $2048^3$ particle 
($4915 h^{-1}$Mpc)$^3$ Cold Dark Matter Simulation (Park et al. 2005$a$).  
We have developed 
a new technique to identify LRG galaxies by finding the most massive 
bound subhalos not subject to tidal disruption (Kim \& Park 2006).  
This $2048^3$ simulation successfully models the 3D topology of the LRG galaxies 
in the Sloan Survey, correctly modeling within approximately $1\sigma$ level
the genus curve found in the observations (Gott et al. 2009).  
By contrast the semi-analytic model of galaxy assignment scheme applied to
the Millennium run by Springel et al. (2005) differed 
from the observational data on topology by $2.5\sigma$ indicating either 
a need for an improvement in its initial conditions (it used a bias factor 
that is too low and an $n_s = 1$ power spectrum that have relatively low 
power at large scales according to WMAP-3 (Spergel et al 2007)) or its galaxy 
formation algorithm (cf. Gott et al. 2008).  The fact that our $2048^3$
N-body simulation modeled well the observational data on LRG 
topology suggests that the formation of LRG galaxies is a relatively clean 
problem that can be well modeled by large N-body simulations using 
WMAP 5-year initial conditions.

Building on this success in modeling the formation and clustering of 
LRG galaxies, in this paper we will show first results from our very 
large Horizon Run N-body simulation in support of the BOSS in the Sloan III survey.


\section{N-Body Simulations}

The first N-body computer 
simulation used 300 particles and was done by Peebles in 1970.  
In 1975, Groth \& Peebles made ``cosmological'' N-body simulations 
using 1,500 particles with $\Omega_m = 1$ and Poisson initial conditions 
(Groth \& Peebles 1975). 
Aarseth, Gott, \& Turner (1979) used 4,000 particles. They found covariance 
functions and multiplicity functions quite like those observed for models with 
$\Omega_m < 1$ and more power on large scales than Poisson (Gott, Turner, \& 
Aarseth 1979; Bhavsar, Gott, \& Aarseth 1981) as originally proposed 
theoretically (Gott \& Rees 1975; Gott \& Turner 1977).  Indeed, the 
inflationary flat-lambda models popular today have $\Omega_m < 1$ and 
more power on large scales than Poisson, just as these early simulations 
suggested. The results were reasonable from theoretical considerations 
of non-linear clustering, concerning cluster (Gunn \& Gott 1972) and void 
(Bertschinger 1985;  Fillmore \& Goldreich 1984) formation from small 
fluctuations via gravitational instability.   The inflationary scenario and 
Cold Dark Matter brought realistic initial conditions for N-body models. 

Geller \& Huchra (1989)'s discovery of the CfA Great Wall of galaxies 
was unexpected.  But no N-body simulations large enough to properly 
model structures as large as the Great Wall had yet been done. 
Then Park (1990), using 4 million particles, a peak biasing scheme, and 
a CDM, $\Omega_m h = 0.2$ model, was able to simulate for the first time 
a volume large enough to properly encompass the CfA Great Wall. A slice 
through this simulation showed a Great Wall just like the CfA Great Wall. 

Large cosmological N-body simulations have been very useful in understanding the
distribution of galaxies in space and time, and provide us a powerful tool
to test galaxy formation mechanisms and cosmological models. The current and future
surveys of large-scale structure in the universe require even larger simulations.
In this paper, we present one of such large simulations to study the distribution 
of LRGs that will be observed by the SDSS-III survey.

\begin{deluxetable*}{c|cccccccccccc}
\tablecaption{Simulation parameters}
\tablewidth{0pt}
\tablehead{
\colhead{$N_p$}
&\colhead{$N_m$}
&\colhead{$L_{box}$}
&\colhead{$N_{step}$}
&\colhead{$z_{i}$}
&\colhead{$h$}
&\colhead{$n$}
&\colhead{$\Omega_m$}
&\colhead{$\Omega_b$}
&\colhead{$\Omega_\Lambda$}
&\colhead{$b$}
&\colhead{$m_p$}
&\colhead{$f_\epsilon$}
}
\startdata
$4120^3$ &$4120^3$&6592 & 400 &23 & 0.72 & 0.96 & 0.26 &0.044&0.74&1.26&$2.96\times 10^{11}$ &160$h^{-1}${\rm kpc}
\enddata
\label{sim}
\tablecomments{
Cols. (1) Number of particles
(2) Size of mesh. Number of initial conditions.
(3) Simulation box size in $h^{-1}$Mpc
(4) Number of steps
(5) Initial redshift
(6) Hubble parameter in 100 km/s/Mpc
(7) Primordial spectral index of $P(k)$
(8) Matter density parameter at $z=0$
(9) Baryon density parameter at $z=0$
(10) Dark energy density parameter at $z=0$
(11) Bias factor
(12) Particle mass in $h^{-1}M_{\odot}$
(13) Gravitational force softening length
}
\end{deluxetable*}

\begin{figure}
\center
\includegraphics[scale=0.45]{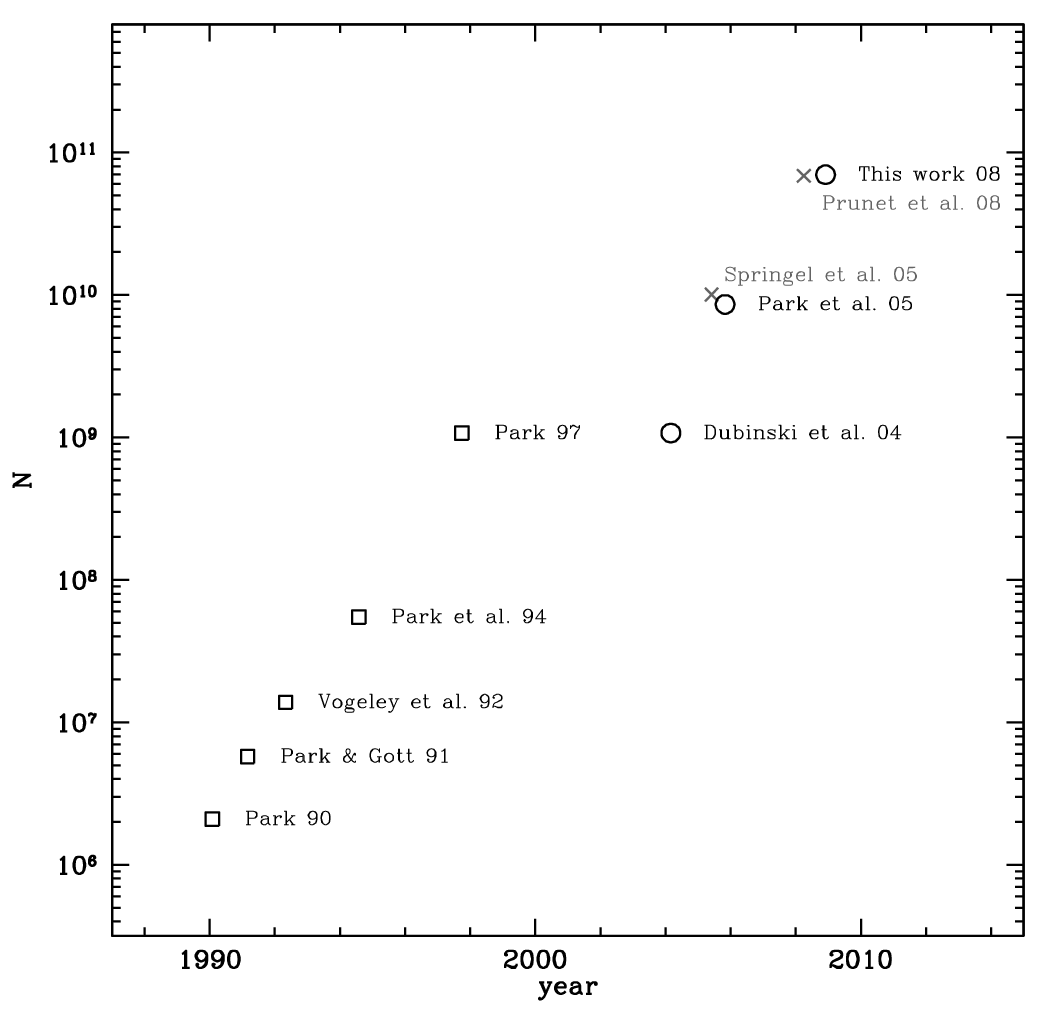}
\caption{
Evolution of number of simulation particles in
the Particle Mesh (PM, squares) and PM-Tree (circles) simulations run 
by the authors, respectively. Crosses are the Millennium and French 
collaboration simulations.
}\label{fig-comp}
\end{figure}

\begin{figure}
\center
\includegraphics[scale=0.38]{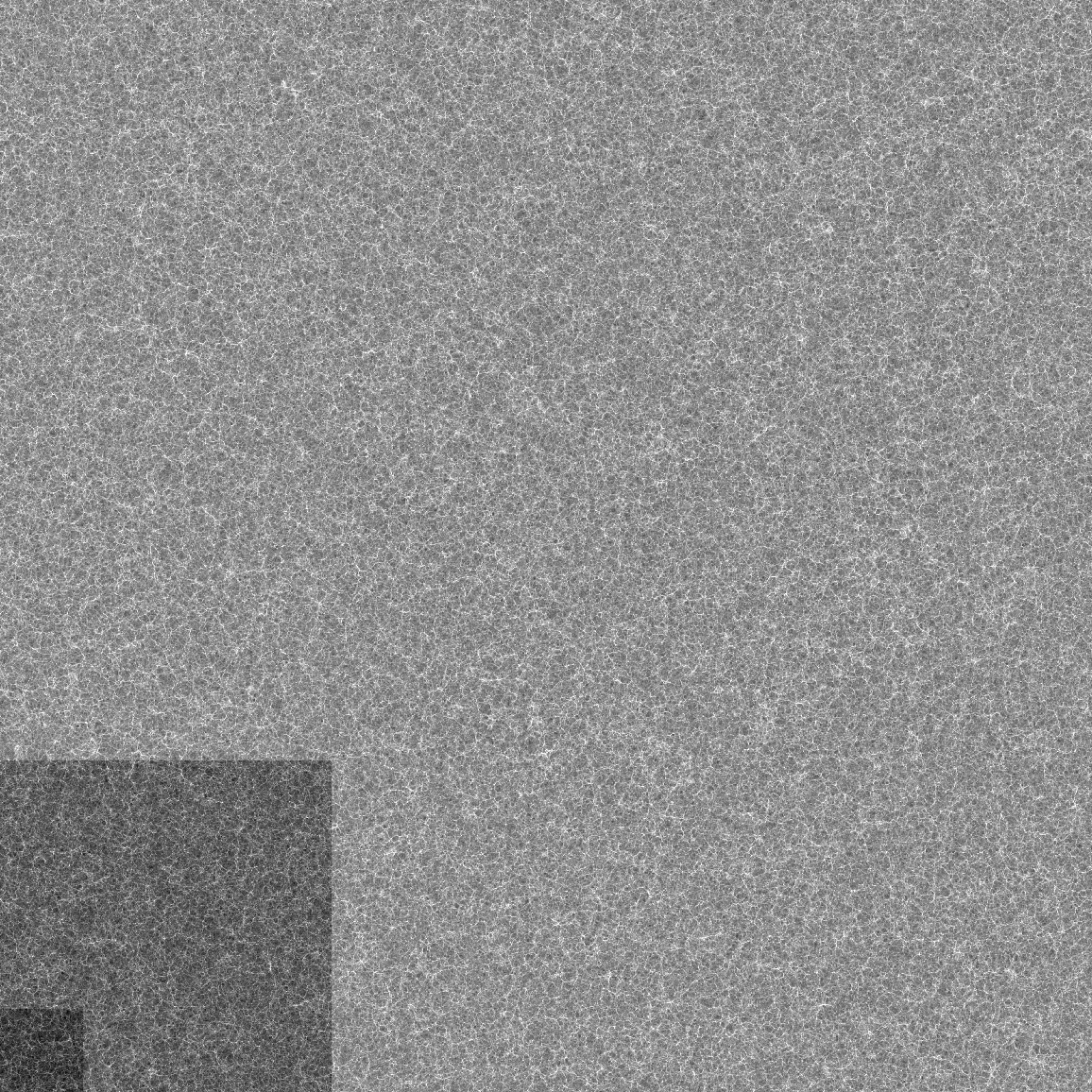}
\caption{
A 4.8$h^{-1}$Mpc-thick slice through our Horizon Run N-body Simulation.
A comparison with the sizes of recent major cosmological N-body simulations
is made.
The side lengths of the simulation cube in our Horizon Run, French
simulation, and the Millennium Run, are 6595, 2000, and 500 $h^{-1}$Mpc,
respectively.
}\label{fig-comp}
\end{figure}

Figure 1 shows how our simulation compares with other recent large simulations 
in terms of number of particles:
the Millennium Run with 10 billion particles, 
and the recent French collaboration N-body simulation (Teyssier et al. 2008) 
using 68.7 billion particles.  Our Horizon Run contains 69.9 billion particles.  
The French simulation size was 
2000 $h^{-1}$Mpc while that of the Millennium run was $500 h^{-1}$ Mpc. 
Figure 2 shows the sizes of these notable simulations to scale for comparison.

\section{The Horizon Run}
Here we describe our Horizon Run N-body Simulation.
We adopt the $\Lambda$CDM model of the universe with the WMAP 5 year
parameters listed in Table 1.
The initial conditions are generated on a $4120^3$ mesh with pixel
size of $1.6 h^{-1}$Mpc. The size of the simulation cube 
is $6592 h^{-1}$Mpc on a side, and $11,418 h^{-1}$Mpc diagonally.
Note that the horizon distance is $10,500 h^{-1}$Mpc.
We use $4120^3=69.9$ billion CDM particles whose initial positions
are perturbed from their uniform distribution
to represent the initial density field at the redshift of $z_i=23$.
They are subsequently gravitationally evolved using a TreePM code
(Dubinski et al. 2004; Park et al. 2005$a$) with a force resolution 
of $160h^{-1}$ kpc, making 400 global time steps taking about 25 days
on the TACHYON\footnote{http://www.ksc.re.kr/eng/resources/resources1.htm},
a SUN Blade system, at KISTI Supercomputing Center in Korea. 
It is a Beowulf system with 188 nodes each of which has 16 CPU cores 
and 32 Gbytes memory.
The simulation used 2.4 TBytes of memory, 20 TBytes of hard disk, and 1648
CPU cores.
We saved the whole cube data at redshifts $z=0, 0.1, 0.3, 0.5, 0.7$, and 1.
We also saved the positions of the simulation particles in a slice 
of constant thickness of $64 h^{-1}$Mpc as they appear in the past light cone.
In order to make the mock SDSS-III LRG surveys we place observers at
8 locations in the simulation cube, and carry out whole sky surveys 
up to $z=0.6$.  These are all past light cone data. 
The effects of our choices of the starting redshift, time step, and
force resolution are discussed in Appendix A, where we use the Zel'dovich
redshifts and FoF halo multiplicity function to estimate the effects.

During the simulation we located 8 equally-spaced (maximally-separated) 
observers in the
simulation cube, and saved the positions and velocities of particles
at $z<0.6$ as they cross the past light cone.
The subhalos are then found in the past light cone data, and used
to simulate the SDSS-III LRG survey.
We assume that the SDSS-III survey will produce a volume-limited LRG sample
with constant number density of $3\times 10^{-4} (h^{-1}Mpc)^{-3}$.
In our simulation we vary the minimum mass limit of subhalos to match
the number density of selected subhalos (the mock LRGs) with this number
at each redshift.
For example, the mass limit yielding LRG number density of 
$3\times 10^{-4} (h^{-1}{\rm Mpc})^{-3}$ 
was found to be $1.33\times10^{13} h^{-1}M_{\odot}$
and $9.75\times10^{12} h^{-1}M_{\odot}$ at $z=0$ and 0.6, respectively.


\begin{figure*}
\center
\includegraphics[scale=0.9]{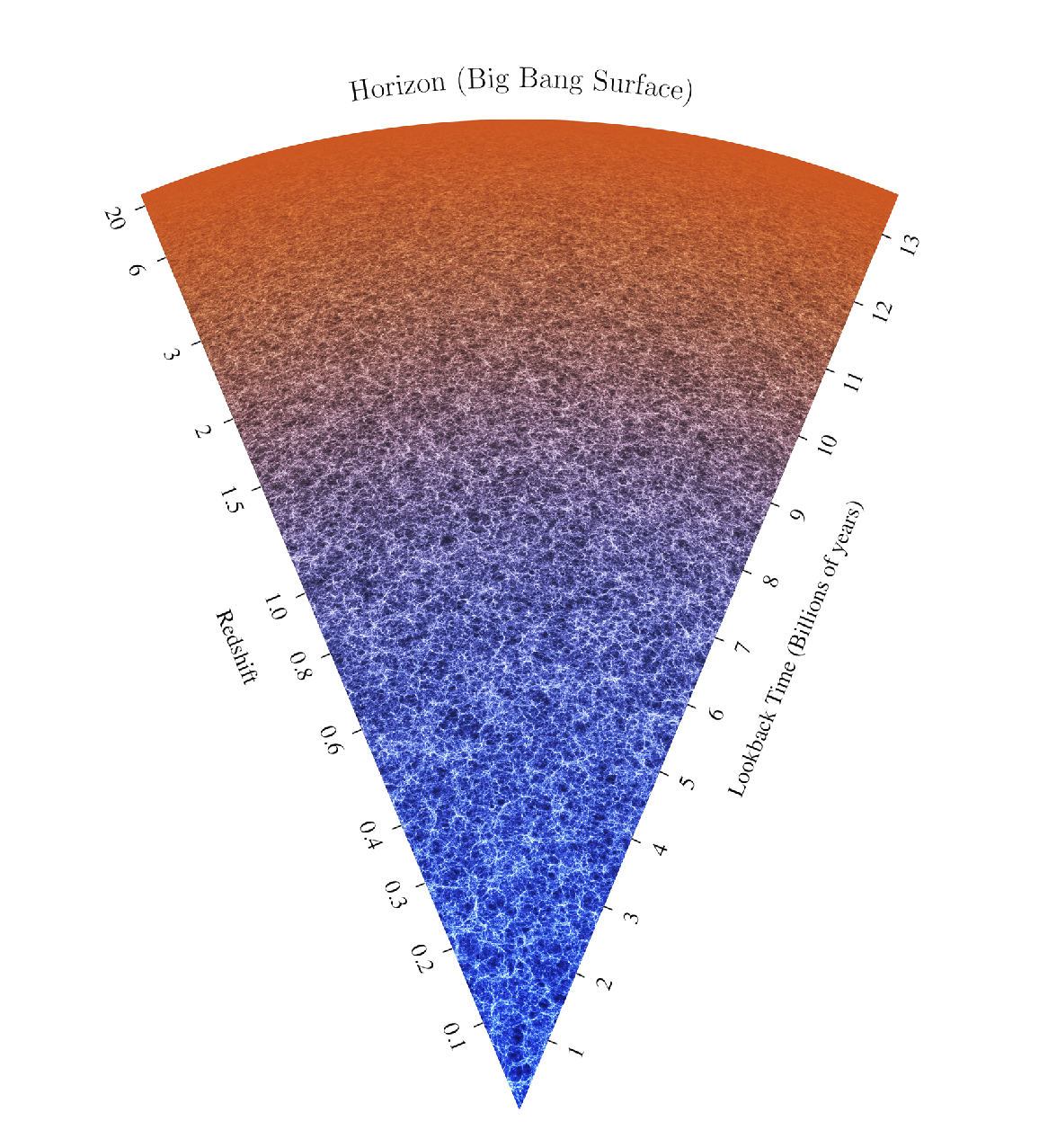}
\caption{
A $64 h^{-1}$ Mpc-thick slice through this simulation showing the matter
density field in the past light cone as a function of look-back time all the way to
the horizon.
The thickness of the wedge is constant and the opening angle is 45 degrees.
The Earth is at the vertex and the upper boundary is the Big Bang surface
at a look-back time of 13.6 billion years.
}
\end{figure*}

Our larger volume allows us to much better model the true power at large scales, 
particularly in the case of the LRG distribution,
which is critical to the baryon oscillation test.  Figure 3 is 
a $64 h^{-1}$ Mpc-thick slice through this simulation showing the matter 
density field in the past light cone all the way to horizon.
The thickness of the wedge is constant and the opening angle is 45 degrees.
The Earth (the observation point)
is at the vertex and the upper boundary is the Big Bang surface
at $z=\infty$. The distribution of the CDM particles is converted to a
density field using the variable-size spline kernel containing 5 CDM particles,
and the density field before $z=23$ was obtained by linearly
evolving the initial density field backward in time.
The radial scale of the slice is the look-back time, so the upper edge
corresponds to the age of the universe, 13.6 billion years located at the
comoving distance of 10,500$h^{-1}$Mpc.

\section{Modelling of the Luminous Red Galaxies}
We use a subhalo finding technique developed by Kim \& Park (2006; 
see also Kim, Park, \& Choi 2008) to identify physically 
self-bound (PSB) Dark Matter subhalos (not tidally disrupted by 
larger structures) at the desired epoch and identify the most massive ones
with LRG galaxies.  This does not throw away information on the 
subhalos which is actually in the N-body simulations and would be thrown away 
in a Halo Occupation Distribution analysis
using just simple friend-of-friend (FoF) halos, and importantly allows 
us to identify the LRG galaxies without free fitting parameters.  
We have a mass resolution high enough to model the formation of LRGs whose
dark halos are more massive than about $M_{30}\simeq 8.87\times10^{12} h^{-1} M_{\odot}$
(total mass of 30 particles).  Within the whole simulation cube
we detect 127,890,474 and 92,043,641 subhalos at $z=0$ and 0.5, respectively.
The mean separation of these massive subhalos is $13.1h^{-1}$Mpc
at $z=0$.

Before we proceed with analyzing the mock LRG sample, it will be interesting
to see how well it reproduces the physical properties of the existing
LRG sample.  Kim et al. (2008) and Gott et al. (2008) has shown that 
the spatial distribution of the subhalos identified in the $\Lambda$CDM simulation 
is consistent with the galaxies in the SDSS Main Galaxy DR4plus sample (Choi et al. 2007).
The observed genus, the distribution of
local density, and morphology-dependent luminosity function
were successfully reproduced by subhalos that are ranked 
according to their mass to match the number density of galaxies.
Moreover, the subhalos identified as LRGs are also shown to reproduce the genus 
topology of SDSS LRG galaxies remarkably well (Gott et al. 2009).
In Appendix B we present a comparison between the correlation functions (CF) of the 
SDSS LRGs and simulated mock LRGs to examine how accurately our mock LRG galaxies 
model the observed ones in terms of the two-point statistic.
We find a reasonable match between them over the scales from $s\approx 1$ to 
140$h^{-1}$Mpc, particularly in the case of the shape of the CF.
Based on our previous and present studies, we conclude that
the observed BAO scale and large-scale topology, which this work is focusing on,
can be very accurately calibrated using the mock galaxies identified 
from our gravity-only simulation. One of the reasons is that both are determined 
by the shape and not by the amplitude of the power spectrum. 

It is interesting to know the fraction of mock LRGs that are central or satellite
subhalos within FoF halos. This can be a test of consistency with
other prescriptions for modeling LRGs (see Zheng et al. 2008 for
the Halo Occupation Distribution modeling).
For comparison with the SDSS LRG sample 
($-23.2 \le {\mathcal M}_{0.3g} \le -21.2$; Zehavi et al. 2005), 
we build 24 mock LRG samples from eight all-sky past light cone data 
with the same observation mask of the SDSS survey.
Because the LRGs brighter than ${\mathcal M}_{0.3g}=-23.2$ 
are extremely rare (Zheng et al. 2008),
we do not apply the upper-mass limit to the mock samples
(refer to Appendix B for detailed descriptions of generating SDSS LRG surveys).
In these mock samples, we count the member galaxies belonging to each FoF halo
utilzing the fact that our PSB method identifies the subhalos or mock LRGs 
within the FoF halos.
Among the mock LRGs in a FoF halo, we regard the most massive LRG 
as the central galaxy and the rest of them as satellites.  
The mass of a host system is obtained by summing up the masses of all member LRGs.
We find that the fraction of the mock LRG galaxies that are satellites in FoF halos is
$5.4\pm0.1$ \%.  This satellite fraction is within the observed range of
5.2 \% -- 6.2 \% (Zheng et al. 2008).

\section{BAO Bump in $\xi(r)$}
\begin{figure}
\center
\includegraphics[scale=0.45]{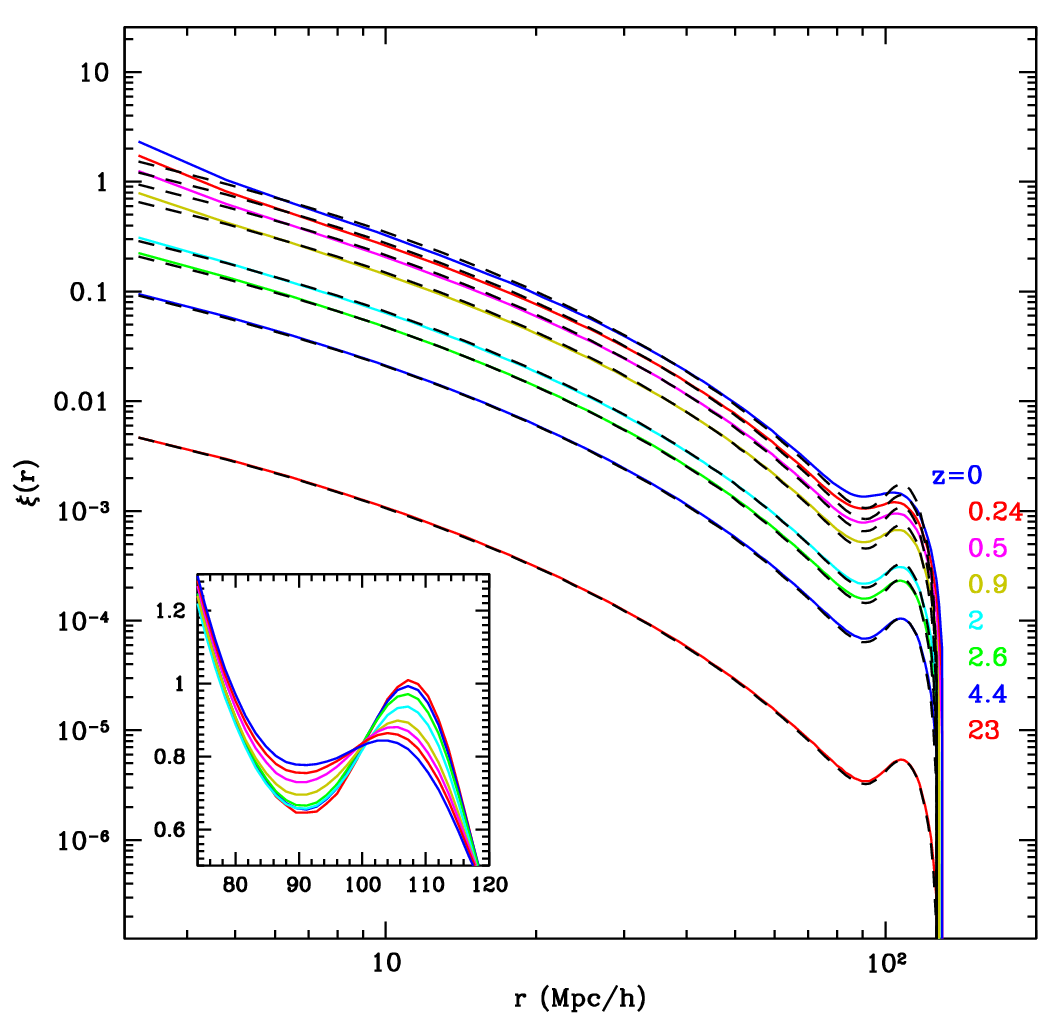}
\caption{
Evolution of the correlation function of
the matter density field at the epoches from $z=0$ to 23.
The dashed curves are the linearly evolved correlation functions, and the colored ones are 
the matter correlation functions measured from the horizon simulation.
The inset box magnifies the matter correlation functions near the baryon 
oscillation bump with amplitudes to match at $r=48 h^{-1}$Mpc after scaling
the peak of the baryonic bump of matter correlation at $z=23$ to unity.
}\label{fig-comp}
\end{figure}

\begin{figure}
\center
\includegraphics[scale=0.45]{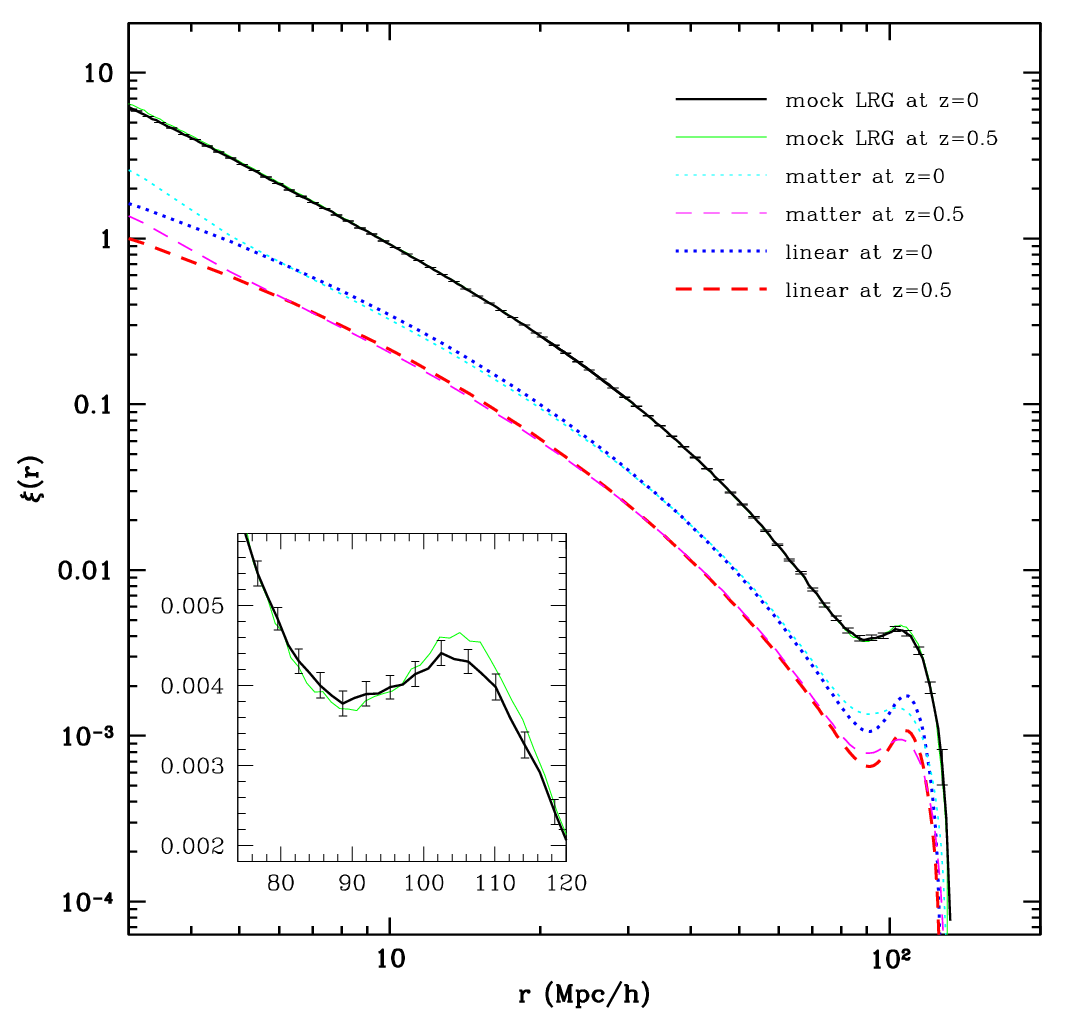}
\caption{
(top curves) The real space correlation functions of the mock LRGs in the whole cube at $z = 0$
and 0.5. $3\sigma$ error bars are attached to the correlation function at $z=0$. 
The matter density correlation functions and the linear theory correlation
functions at $z=0$ and 0.5 (bottom curves), are also shown.
}\label{fig-comp}
\end{figure}

Figure 4 shows the evolution of the CF of
the matter density field from $z=23$ to $z=0$. The dashed curves are
the linear CFs, and colored ones are the matter CFs measured at 
eight redshifts using the whole simulation data. 
The inset box demonstrates the change in the shape of CF near 
the baryon oscillation bump due to the non-linear gravitational evolution
in the matter field. 
In the inset box the amplitude of the CF at $z=23$ is scaled so that
the peak of the baryon bump has unit amplitude, and then the amplitudes 
of the CFs at other epoches are scaled to match the CF at $z=23$ at 
$r=48h^{-1}$Mpc.  The amplitude of the bump decreases by 16\%,
and the peak location decreases by 4\% from
the initial epoch to the present.

The top two curves of Figure 5 are the CFs calculated from 
all mock LRGs in the whole cube at $z = 0$ 
and 0.5. They are also shown in the inset box at the lower left corner.
The curve with $1\sigma$ error bars is the mock LRG CF at $z=0$.
Errors are estimated from sub-cube results.
The baryon oscillation bump is clearly visible.  
Notice the tiny amount of noise, because we are sampling such a large volume.   
In Figure 5 we also present the
3D covariance function of the matter density field, and 
linear regime CF at $z=0$ and 0.5 for comparison.  
The linear theory gives $107.6  h^{-1}$Mpc for the
position of the BAO peak while in the mock LRG CFs they
are  $103.6 h^{-1}$Mpc and $104.7  h^{-1}$Mpc 
at $z =0$ and 0.5, a difference of up to 3.7 \% and 2.7 \%, respectively. 
On the other hand, the matter density field has peak at 
$103.2 h^{-1}$Mpc and $105.0  h^{-1}$Mpc at these epoches, respectively.

It is important to note that the systematic effects in the baryonic bump
in the LRG CF depend on the type of tracer.
Because of the high statistical biasing in the distribution of LRGs
the evolution of the LRG CF is different from that of the matter CF.
In Figure 6 we show the ratio of the matter and mock
LRG CFs relative to the linear CF.
The bottom three curves are the matter CFs at $z=0, 0.5,$ and 2.6
divided by the linear CF evolved to the corresponding redshifts.
The upper two solid curves are the mock LRG CFs at $z=0$ and 0.5 also 
divided by the linear CF evolved to  $z=0$ and 0.5.
It can be noted that the deviation from the shape of the linear CF is 
different in the cases of the matter and mock LRG CFs
even though the difference is not large.
For example, at the separations of $r=48$ and 70 $h^{-1}$ Mpc the ratio between
the LRGs and matter at $z=0.5$ is about 4.65, but $r=105h^{-1}$Mpc the ratio 
is 4.52 a 2.7\% difference.
The long dashed lines in Figure 6 are the ratios of the mock LRG CF to
the evolved matter CF.
We can observe small enhancements of the ratio
around the baryonic acoustic signature both at $z=0$ and 0.5,
and this enhancement is consistent with the results of Desjacques (2008) 
who used density maxima instead of halos.
Therefore, it will be erroneous to use the non-linear matter CF
to correct for the observed LRG CF.
It is very important to correctly model the LRG galaxies in the
$N$-body simulation that have the accurate degree of biasing, and
to use them for the correction.

\begin{figure}
\plotone{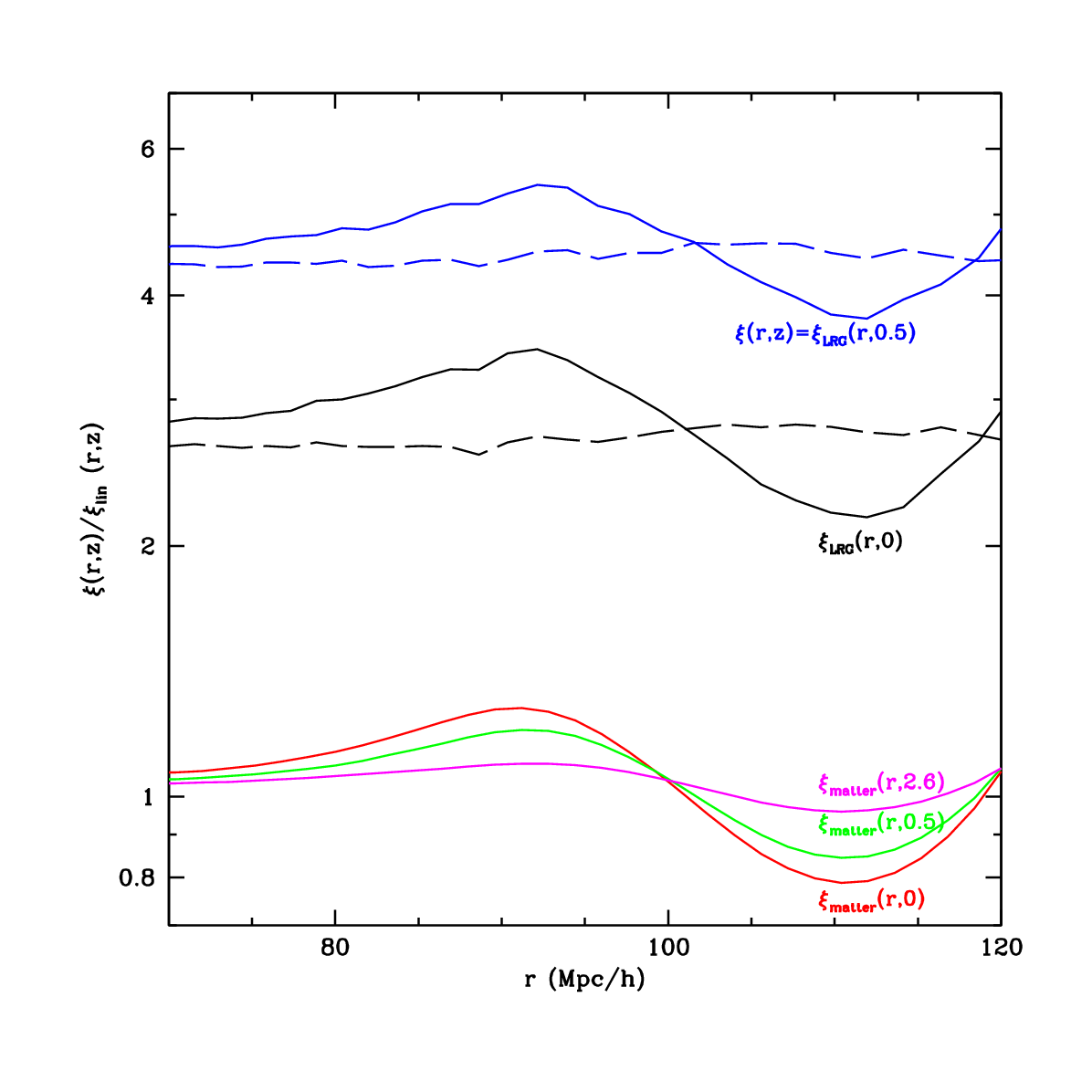}
\caption{
Ratio of the matter and mock LRG correlation functions 
relative to the linear correlation function. Upper two solid curves 
are for mock LRGs at $z=0$ and 0.5. Their correlation functions are
divided by the linear CFs evolved to  $z=0$ and 0.5.
Two dashed curves are the ratios of correlation functions
of the mock LRGs to those of matter fields.
Bottom three curves are for matter density fields at $z=0, 0.5,$ and 2.6.
Their correlation functions are divided by the linear CFs evolved to 
the corresponding redshifts.
}
\end{figure}

\begin{figure*}
\plotone{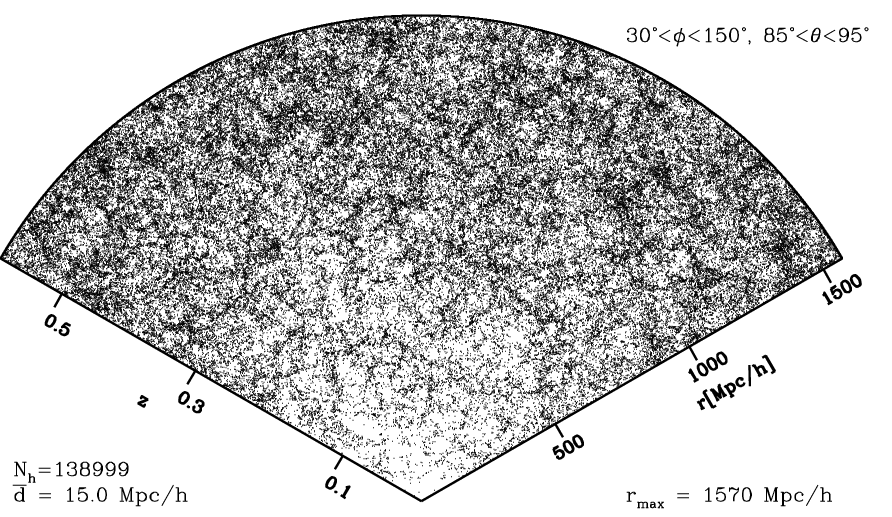}
\caption{
A 10 degree-thick wedge diagram of a slice taken from a mock 
LRG survey performed  at a random location observed
along the past light cone of the Horizon Run.
The mean separation between the LRG galaxies is set to 15 $h^{-1}$Mpc
at all redshifts, and the outer boundary of the survey is at $z=0.6$.
}
\end{figure*}

For direct comparison with the Sloan III data, we have made all-sky
LRG galaxy surveys using the past light cone data
generated at 8 locations in our Horizon Run. The
surveys are made so as to maintain the mean number density of LRGs 
as $3\times 10^{-4} (h^{-1}{\rm Mpc})^{-3}$.  Figure 7 shows a fan 
shaped diagram of a random slice through a mock Sloan III survey.  

In Figure 8 we present the CFs of the LRGs observed in the mock SDSS-III 
surveys.
At each observer location we make four SDSS-III LRG survey catalogs
covering $\pi$ stradian of the sky and extending out to $z=0.6$.
Each mock LRG sample is then divided into three redshift shells that
contain roughly equal number of galaxies. The redshift boundaries between
these shells are at $z_1=0.396$ and $z_2=0.513$.
Three solid lines of Figure 8 are the mean CFs averaged over 32 mock surveys
measured in these three shells. We add peculiar velocities along
the line-of-sight to simulate the effects of redshift distortion, 
and measure the covariance function in the past light cone.
The $1\sigma$ error bars for the CF of each redshift shell, 
obtained from 32 mock surveys, are 
attached to the CF of the lowest redshift shell ($R_1: 0<z<z_1$).
We observe in the redshift-distorted samples a substantial increase 
in the CF amplitude on larger scales ($s\gtrsim 40 h^{-1}{\rm Mpc}$),
especially on the BAO scale. 
This shift of the BAO amplitude
between redshift slices was also seen in the Fig. 19 of Cabre \& Gaztanaga (2009)
who used SDSS LRG samples for their analysis.
They tried to explain the BAO-amplitude shift by introducing possible systematic 
effects in the SDSS observation but they did not give any conclusion. 
However, it is interesting to observe a similar shift in the mock surveys which
does not suffer from the systematic effect.

The local peak in the CF measured from individual mock survey data
shows a large variation, and so we developed a method that can measure the
BAO bump position more accurately.
The position of the BAO bump was measured by fitting the observed CF to 
a set of template CFs.  The template CFs are those selected from 
the simulated matter CFs from $z=23$ to 0. 
As shown in Figure 4, the matter CF is accurately known, and
shows a wide range of variation in the amplitude and shift of the BAO bump.
To fit an observed CF to the templates, we calculate
%
\begin{equation}
\chi_m^2 =\sum_{r_1 \le r\le r_2} \left[\xi(r) - A_m\xi_m(r-\delta_m)\right]^2
\end{equation}
where $A_m$ and $\delta_m$  are the amplitude and shift parameters.
Here we set $r_1=90$ and $r_2=115 h^{-1}{\rm Mpc}$ to safely include the bump.
We numerically find the free parameters $A_m$ and $\delta_m$ 
that minimize $\chi^2$ to determine the best-fit template, 
and to estimate the position of the BAO bump.

A close inspection reveals 
that the peak location and amplitude of the BAO bump in the LRG CF
slowly decrease as redshift decreases; the location of the BAO bump
is at $102.0\pm 5.0$, $102.8\pm 3.9$, and $103.8\pm 4.1$ $h^{-1}$Mpc 
for $R_1(0<z<z_1), R_2 (z_1 < z < z_2)$, and $R_3 (z_2 < z < 0.6)$ shells, respectively.
Due to the sample variance the uncertainty in the BAO scale
amounts as much as 5.0\%.
For a comparison the real space and redshift space CFs 
of all LRGs in the simulation box at $z=0.5$
are also plotted in dotted and dashed curves, respectively. Their BAO bump peaks are
located at 105.2 and $102.5 h^{-1}$Mpc, respectively.

The uncertainty in the BAO bump position can be reduced if larger samples are used.
We measure the CF of mock LRG galaxies
located in a redshift shell with inner and outer boundaries at $z=0.4$ and 0.6,
respectively. This region contains more 
data than previous settings leading to smaller scatter 
around the average (see Fig. 9).
The mean value of the BAO peak position over 32 mock surveys is 103.2$h^{-1}$Mpc and
the difference from the linear regime initial conditions is
$4.4(\pm 2.7) h^{-1}$Mpc.
Thus if the BAO peak is measured as in our analysis, 
the peak location should be corrected upward by above value.  
The uncertainty 
in this systematic correction is also estimated from our 32 mock 
SDSS-III surveys.
%
%
This gives us a benchmark idea of the extent of 
the systematic effects (and their uncertainty) encountered in measuring 
the position of the BAO bump from LRGs and comparing it with the baryon 
oscillation scale from the linear regime theory.  
As hoped for, the systematic effects 
due to non-linear effects and biasing are small, and can be corrected 
for with high accuracy.  
The corrected value will only be in error by the uncertainty in the correction 
factor which is 2.5\% when the data in this thick shell is used.   
%
Comparison of the individual 32 mock 
Sloan III surveys after such systematic corrections have been made will 
give us an accurate measure of the statistical accuracy of the results on 
$w(z)$ that we should expect in the real survey.

\begin{figure}
\center
\includegraphics[scale=0.45]{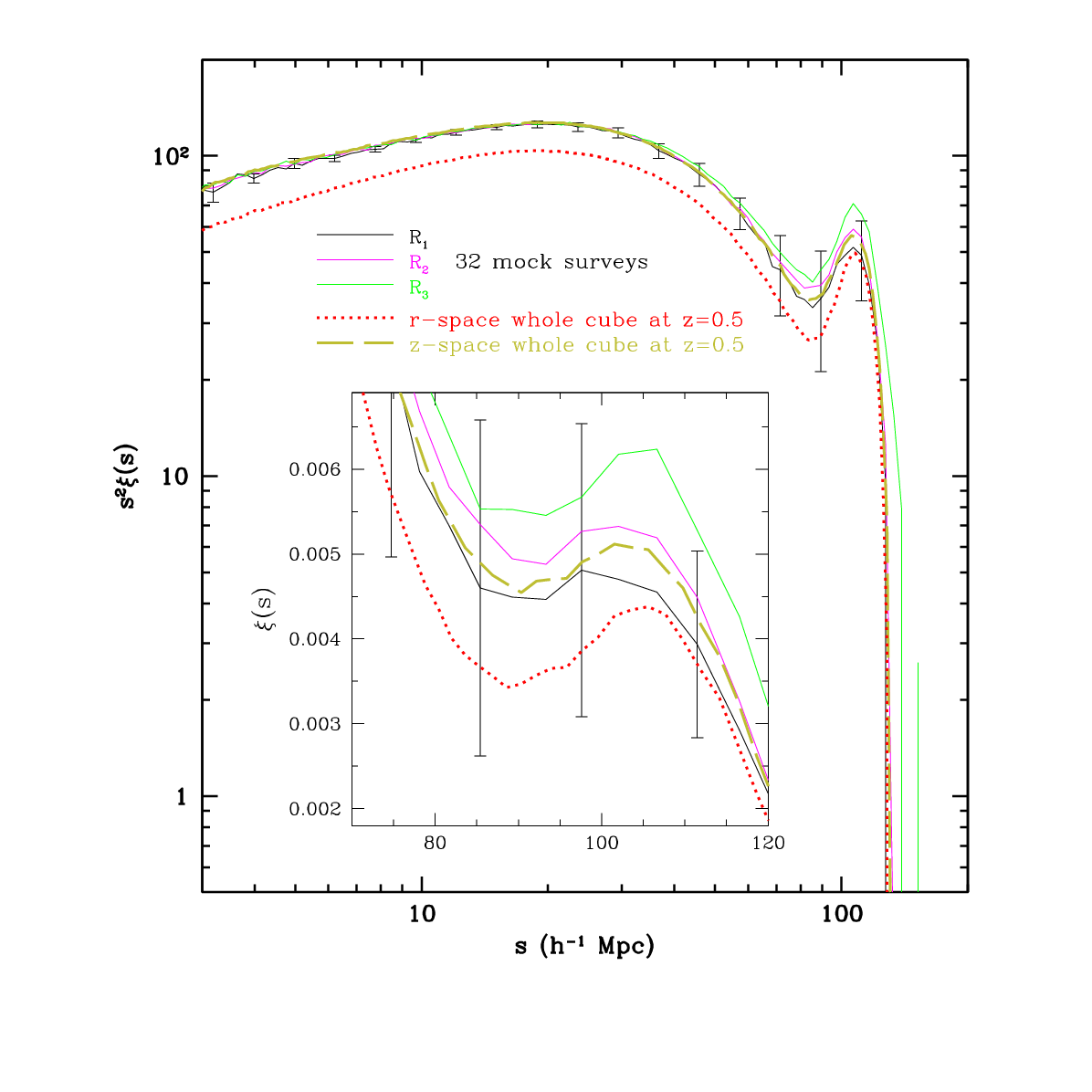}
\caption{({\it Solid lines}) Mean correlation functions of the LRGs 
averaged over 32 mock SDSS-III surveys. 
Each mock survey is divided into three shells with redshift boundaries at
 $z_1=0.396$ and $z_2=0.513$. 
$1\sigma$ error bars for each mock survey are attached to the CF of
the shallow sample $R_1$. ({\it Dotted line}) Correlation function of the mock LRGs
in the whole simulation cube at $z=0.5$. ({\it Dashed line}) The linear
theory correlation function evolved to the present epoch.
}\label{fig-comp}
\end{figure}


\begin{figure}
\center
\includegraphics[scale=0.45]{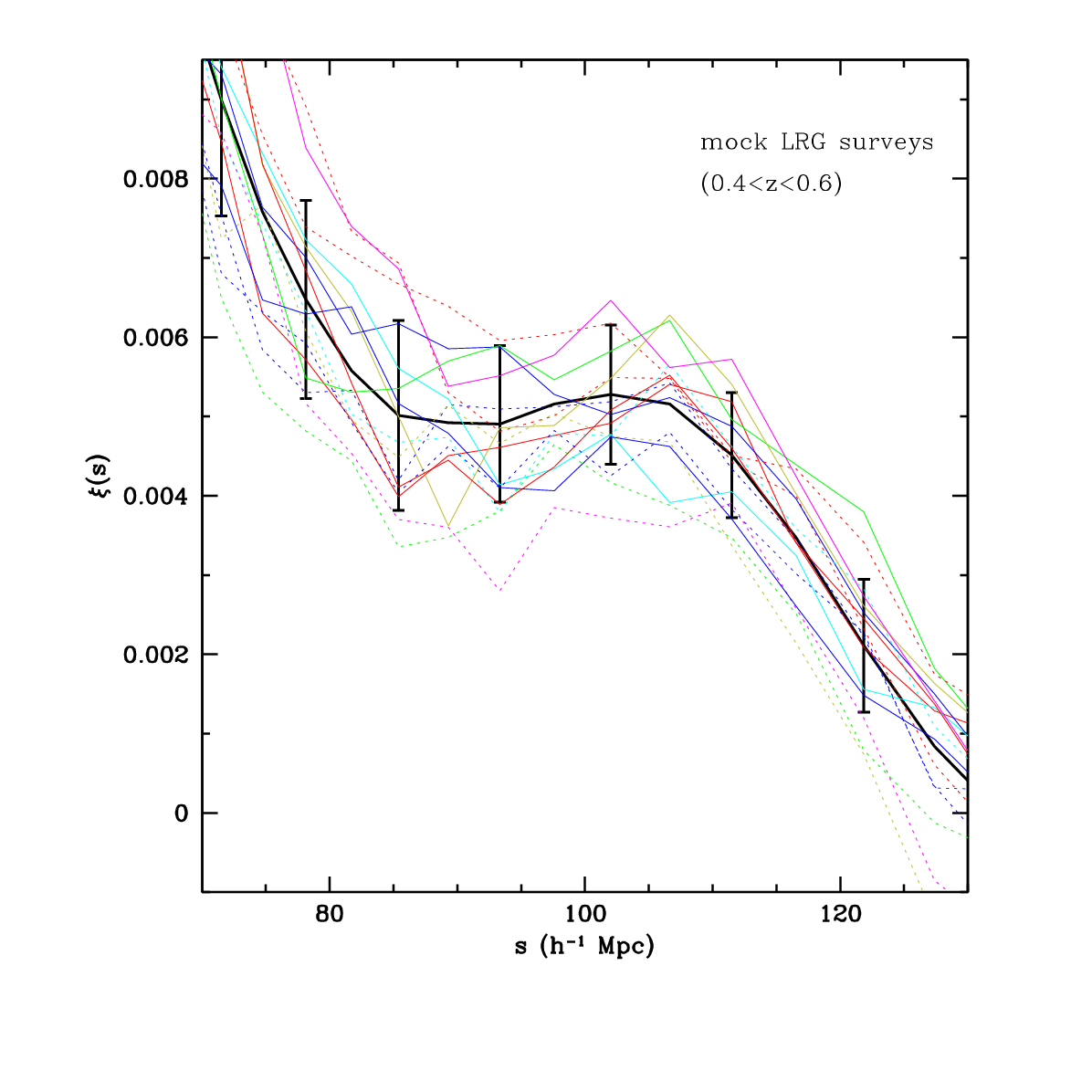}
\caption{({\it Colored lines}) Correlation functions
of 16 mock SDSS-III surveys and their mean ({\it black lines with error bars})
in a $0.4 < z < 0.6$ redshift shell.
For clarity, we show only half the mock survey results.
The position of the baryon oscillation bump is measured by fitting 
the correlation function of the mock LRGs to a set of
template correlation functions.
}\label{fig-comp}
\end{figure}

These preliminary studies can be 
used to improve the survey strategy and reconstruction methods
before the survey starts. 
In the simulations we are omniscient since we know where each of the LRG 
galaxies actually originated, so this should help us to design and improve 
the reconstruction techniques.  As the data is being collected, and 
the selection function of the LRG galaxies being detected becomes accurately 
known, it will be possible to improve the accuracy of these studies 
as the survey continues. Comparisons with the N-body simulations should be 
of critical value, allowing one to calibrate the dark energy experiment 
with high accuracy.

\section{BAO Wiggles in $P(k)$}
A BAO bump in the correlation has a wiggle pattern in the power spectrum (hereafter PS).
Also similar changes, the shift and degrading of shapes,
happen to the baryonic wiggles of PS.
Recently, Seo et al. (2008) reported that they have detected sub percentage level
of shift of the BAO wiggle 
after applying a fitting to the matter PS. Therefore, it is valuable to check 
whether LRG galaxies, as a biased peak,
have a similar level of shift or they have different scale of shift
compared to the matter field.
Figure 10 shows the mock LRG power spectra
in real (the second top line) and redshift (topmost line) 
spaces at $z=0$ and matter PS at $z=0.5$ (bottom line) and 0.
Figure 11 shows the evolution of the simulated matter PS
normalized by the smoothed PS ($P_{sm}$) 
around baryonic scales (Eisenstein \& Hu 1998).
As can be seen, the non-linear gravitational evolution
affects the wiggle pattern from small scales
making them degraded in shape and buried in nonlinear background with time.
Oscillatory features at $k \ge 0.1 h/$Mpc
experience a severe distortion in shape while the
features on the larger scales show less deviations from the linear shapes.

\begin{figure}
\center
\includegraphics[scale=0.45]{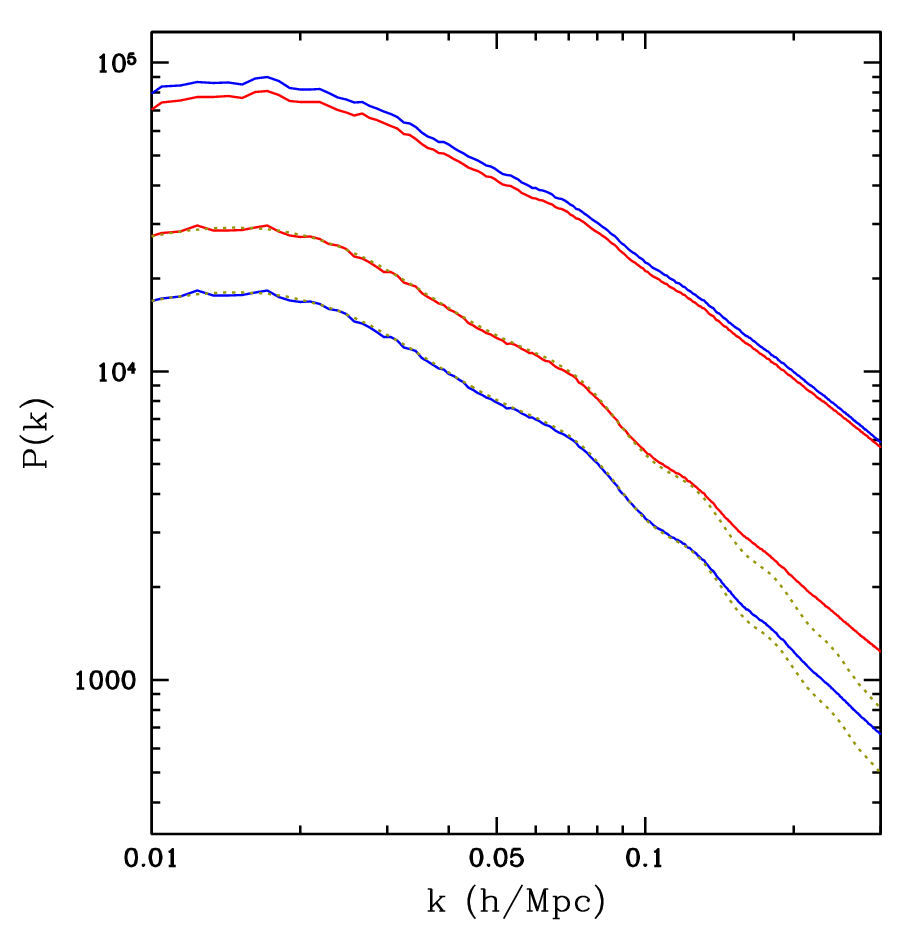}
\caption{
Power spectra of matter fields at $z=0.5$ (two bottom lines)
and at $z=0$ (two middle lines). 
Among matter power spectra, 
the solid lines are those from the simulation and the dotted lines
are the linearly evolved ones.
Top curves are those for the LRG galaxies
in redshift (topmost curve) and real spaces at $z=0$.
}\label{fig-pksm}
\end{figure}

\begin{figure}
\center
\includegraphics[scale=0.45]{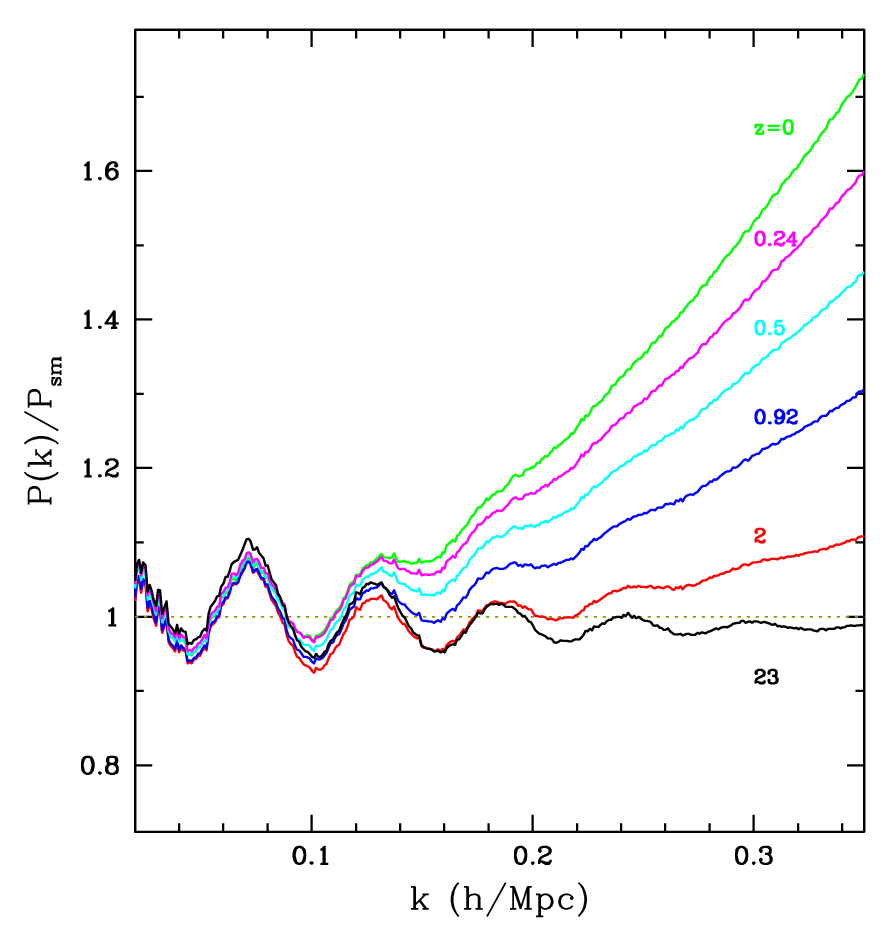}
\caption{
The ratio of the non-linear matter power spectrum to the smoothed 
linear power spectrum at redshifts from $z=$ 23 to 0. 
}\label{fig-pksm2}
\end{figure}

To quantify the non-linear distortion of the BAO feature in the PS
of mock LRG galaxies, we followed the analysis of Seo et al. (2008)
who introduced the shift ($\alpha$) and degrading ($\Sigma_m$) parameters
in the polynomial fitting functions.
The contrast of the baryonic feature 
is obtained by subtracting 
the smoothed background PS ($P_{sm}$) of no wiggle form
from the linear PS ($P_{lin}$): 
\begin{equation}
P_b(k) \equiv P_{lin}(k) - P_{sm}(k).
\end{equation}
The fitting function is organized as
\begin{equation}
P_{fit}(k) = B(k) P_m(k/\alpha) + A(k),
\label{fiteq}
\end{equation}
where $B(k)$ reflects the scale-dependent biasing and mode coupling,
and $A(k)$ is added to account for shot noise.
For the $\chi^2$-fit, we adopt the polynomial forms:
\begin{eqnarray}
\nonumber
A(k)&=& \sum_{i=0}^7 a_i k^i \\
B(k)&=& \sum_{i=0}^2 b_i k^i.
\end{eqnarray}
This functional form provides a reasonable fit to the data and, therefore, we 
will not consider other alternatives listed in Seo et al. (2008).
And $P_m(k/\alpha)$ is the shifted model PS with a degrading
factor:
\begin{equation}
P_m(k) = P_b(k) \exp{\left(-k^2\Sigma_m^2/2\right)}
+ P_{sm}(k).
\end{equation}
The linear PS subtracted by the smoothed
power is suppressed by a Gaussian damping of length, $\Sigma_m$.
Seo et al. (2008) fixed it as $\Sigma_m=7.6 h^{-1}$Mpc for the matter field at $z=0.3$.
However, we treat it as a free parameter here because 
we are dealing with biased subhalos while Seo et al. used the matter particles.
The shift parameter, $\alpha$, is added to account for the shift of the
baryonic features due to the non-linear evolution and biasing.

\begin{figure}
\center
\includegraphics[scale=0.55]{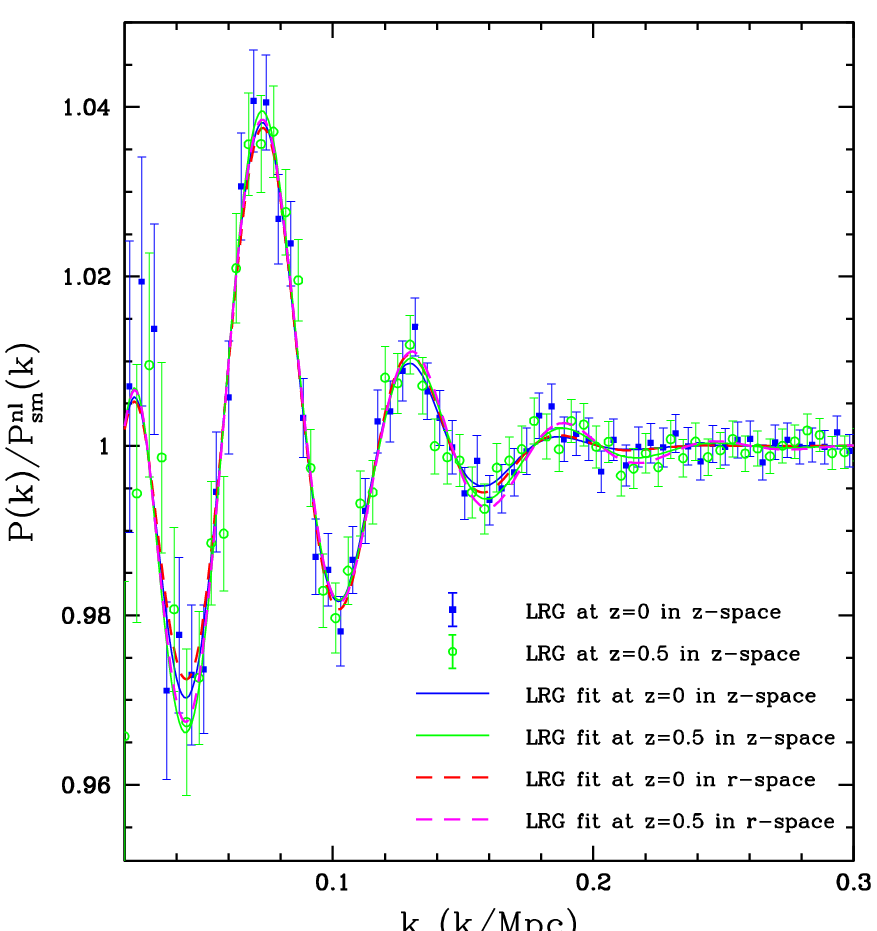}
\caption{
Baryonic features of LRG galaxies.  The ratio of the power
spectrum of LRG galaxies to the smoothed non-linear 
power spectrum at $z=0$ and $z=0.5$.
Filled boxes and open circles show the measured values of LRG
galaxies in redshift space
at $z=0$ and $z=0.5$, respectively.
}\label{fig-bar}
\end{figure}
In Figure \ref{fig-bar} shown are the ratio of the LRG PS to
the smoothed non-linear PS ($P^{nl}_{sm}(k)$).
It shows the contrast of the BAO wiggles over the smooth background PS.
The smoothed PS in the non-linear regime is obtained by applying
\begin{equation}
P^{nl}_{sm} \equiv B(k) P_{sm}(k/\alpha) + A(k),
\end{equation}
where we adopt the same values of $A(k)$ and $B(k)$ as 
previously obtained when performing the fitting to the simulated power
spectrum of the LRG galaxies.

The fitting results are $(1-\alpha)=-0.020$, $\Sigma_m=14.9 h^{-1}$Mpc,
and $\chi^2/$DOF=0.43 at $z=0$ and,
$(1-\alpha)=-0.023$, $\Sigma_m=9.0 h^{-1}$Mpc,
and $\chi^2/$DOF=0.47 at $z=0.5$ for LRG galaxies in redshift space.
For the real-space data, $\alpha$ is unchanged while $\Sigma_m$ decreases slightly.
The negative value of $1-\alpha$ means that the baryonic feature
of the LRG galaxies is moving toward small scales relative to the linear PS.
We detected a 2\% wiggle shift for the LRG galaxies
while Seo et al. (2008) found 0.45\% at $z=0.3$ for the matter field.
This significantly larger shift is mainly due to the biasing effect 
and should be kept in mind when interpreting the observational data 
to recover the oscillation scales in the PS.
Also the degrading parameter ($\Sigma_m=9.0h^{-1}$Mpc)
is larger than $\Sigma_m=7.6 h^{-1}$Mpc at $z=0.3$
which was adopted for matter field by Seo et al. (2008). 
The more negative shift and larger degrading parameter 
for the mock LRG galaxies relative to the background matter density field,
again demonstrate the tracer dependence of the non-linear effects.
We have also applied above analysis to the matter field in real space.
The value of $1-\alpha$ is $-0.0025, -0.0025, -0.001$, and 0 at $z=0$, 0.5, 2, 
and 23, respectively.
Also we obtain $\Sigma_m=8.25$, 6.60, 3.75, and 0 $h^{-1}$Mpc for each redshift.
These results are consistent with Seo et al. (2008).

\section{Genus Topology of Large Scale Distribution of LRG Galaxies}
Through the genus topology analysis
one can check whether or not the mock LRG galaxies in
our simulation correctly reproduce the spatial distribution of
the observed LRGs.  
At scales larger than the correlation length the LRG genus curve
is well-approximated by the Gaussian random phase curve with 
expected small shifts due to non-linear effects and biasing.   

We have developed tools for analyzing the genus topology of large scale 
structure in the universe (Gott, Dickinson, \& Melott 1986; Hamilton, 
Gott, \& Weinberg 1986; Gott, Weinberg, \& Melott 1987; Gott, et. al. 1989; 
Vogeley, et. al. 1994; Park, Kim, \& Gott 2005$a$; Park et. al. 2005$b$). 
We smooth the spatial distribution of galaxies to construct density 
contour surfaces.  We measure 
the genus as a function of density, allowing comparison with the topology 
expected for Gaussian random phase initial conditions, as predicted 
in a standard big bang inflationary model (Guth 1981; Linde 1983) where 
structure originates from random quantum fluctuations in the early universe. 
We smooth the galaxy distribution with a Gaussian smoothing ball 
of radius $R_G$, where $R_G$ is 
chosen to be greater than or equal to the correlation length to reduce 
the effects of non-linearity, and greater than or equal to 
the mean particle separation to reduce the shot noise effects.  

     Density contour surfaces are labeled by $\nu$, where the volume 
fraction on the high density side of the density contour surface is $f$:
\begin{equation}
f = {1\over \sqrt{2\pi}} \int_{\nu}^{\infty} {\rm exp}(-x^2/2) dx.
\end{equation}

The genus curve is given by the topology of the density contour surface:
\begin{equation}
g(\nu) = {\rm  \#~of~ donut~ holes} - {\rm \#~ of~ isolated~ regions}
\end{equation}
(Gott et al. 1986).  An isolated cluster has a genus of  
$-1$ by this definition.  Since $g(\nu)$ is equal to minus the integral of 
the Gaussian curvature over the area of the contour surface divided 
by $4\pi$, we can measure the genus with a computer program (CONTOUR3D) 
(see Gott et al. 1986, 1987).
For Gaussian random phase initial conditions:
\begin{equation}
g(\nu) = A(1-\nu^2) {\rm exp}(-\nu^2/2),
\end{equation}
where $A = (\langle k^2\rangle/3)^{3/2}/(2\pi)^2$ and 
$\langle k^2\rangle$ is the average value of 
$k^2$ integrated over the smoothed power spectrum (Hamilton et al.
1986; Adler 1981; Doroshkevich 1970; Gott et al. 1987).  
The $f = 50$\% median density contour ($\nu = 0$) shows a predicted 
sponge-like topology (holes), the $f = 7$\% high density contour ($\nu = 1.29$) 
shows isolated clusters while the $f = 93$\% density contour   ($\nu = -1.29$) 
shows isolated voids.

When the smoothing length is much greater than the correlation length, 
fluctuations are still in the linear regime and since fluctuations in 
the linear regime grow in place without changing topology, the topology 
we measure now should reflect that of the initial conditions, which 
should be Gaussian random phase according to the theory of inflation 
(Gott et al. 1987; Melott 1987). Small deviations from the random phase 
curve give important information about biased galaxy formation and 
non-linear gravitational effects as shown by perturbation theories 
and large N-body simulations 
(c.f. Matsubara 1994; Park et al. 2005$a$).  All previous studies 
have shown a sponge-like median density contour as expected from 
inflation (Gott et al. 1986, 1989, 2009;
Moore et al. 1992; Vogeley et al. 1994; Canavezes et al. 1998; 
Hikage et al. 2002, 2003; Park et al. 2005$b$)

Small deviations from the Gaussian random phase distribution are expected 
because of non-linear gravitational evolution and biased galaxy formation 
and these can now be observed with sufficient accuracy to do model 
testing of galaxy formation scenarios.

The genus curve can be parameterized by several variables.  The amplitude $A$ 
of the genus curve, proportional to $\langle k^2\rangle^{3/2}$ 
of the smoothed power 
spectrum, gives information about the power spectrum and on
the phase correlation of the density distribution.
Shifts and deviations in the genus curve from the overall theoretical 
random phase case can be quantified by the following variables: 
\begin{equation}
\Delta\nu = \int_{-1}^{1} g(\nu)\nu d\nu /  
\int_{-1}^{1}g_{\rm rf}(\nu)d\nu,
\end{equation}
\begin{eqnarray}
A_V &=& \int_{-2.2}^{-1.2}g(\nu)d\nu /  \int_{-2.2}^{-1.2}g_{\rm rf}(\nu)d\nu,\\
A_C &=& \int_{1.2}^{2.2}g(\nu)d\nu /  \int_{1.2}^{2.2}g_{\rm rf}(\nu)d\nu,
\end{eqnarray}
where $g_{\rm rf}(\nu)$ is the genus of the random phase curve 
(Eq. 8) best fit to the data. 
Thus, $\Delta\nu$ measures any shift in the central portion of 
the genus curve.  The theoretical curve (Eq. 8) has $\Delta\nu = 0$.  
A negative value of $\Delta\nu$ is called a ``meatball shift'' caused 
by a greater prominence of isolated connected high-density structures 
which push the genus curve to the left. This can be due to non-linear 
galaxy clustering and bias associated with galaxy formation.  $A_V$  
(and $A_C$) measure the observed number of voids (and clusters), respectively, 
relative to those expected from the best fitting theoretical curve. 

Park et al. (2005$a$) show that $A_V < 1$ can result from biasing 
in galaxy formation because voids are very empty and can coalesce 
into a few larger voids. A value of $A_C < 1$ can occur because of 
non-linear clustering, when clusters collide and merge, and if there 
is a single large connected structure like the Sloan Great Wall.  
As Park et al. (2005$a$) have shown, with Matsubara's (1994) formula 
for second-order gravitational non-linear effects alone, one has the 
result that $A_V + A_C = 2$ at all scales, so if we observe both 
$A_V$ and $A_C$ to be less than 1, for example, 
biased galaxy formation must be involved.  

Gott et al. (2009) compared 
observational genus curves from the SDSS survey with a few recent 
$N$-body simulations.  
The semi-analytic model of galaxy assignment scheme applied to the 
Millennium run (Springel et al 2005) gives reasonable values 
for $A_V$ and $A_C$, but 
its value of $\Delta\nu$ differs from the observational data by 2.5 
$\sigma$, showing less of a ``meatball'' shift than the data.
This indicates a need 
for an improvement in either its initial conditions (it used a bias factor 
that is too low according to WMAP 5 year results) or 
its galaxy formation algorithm (Gott et al. 2008).   

Gott et al. (2009) recently measured
the 3D topology of the LRG galaxies from the Sloan Survey 
and measured small deviations from the Gaussian form using $\Delta\nu,
A_V$, and $A_C$ from Equation 10 to 12.  There is only 
a small amount of noise in the genus curve due to the large sample size.  
The results are compared with the genus curves averaged over
twelve mock LRG galaxy surveys that are performed in
a $2048^3$ particle $\Lambda$CDM simulation (Gott et al. 2009).
The mean of the mock surveys fits the observations extraordinarily well, 
including the fact that the number of voids, as measured by the depth 
of the valley at $\nu  < -1$ or by $A_V$, is less than the number of 
clusters, as measured by the depth of the valley at $\nu > 1$ or
by $A_C$, which is also seen in the observations.
The fact that the clusters outnumber the voids is due to 
non-linear biasing and gravitational evolution effects, 
and they are modeled by the simulations very well.  
The observed amplitude of the genus curve $A$ at $R_G=21$ and $34 h^{-1}$Mpc scales 
agrees extremely well with the mean of the mock N-body catalog. This indicates that
the shape of power spectrum is also modeled extremely well.   
The observed values of $A, \Delta\nu, A_V$, 
and $A_C$ are within approximately $1\sigma$ of those expected \
from the 12 mock catalogs, all without any free fitting parameters
(see Gott et al. 2009 for details).
This suggests that finding the LRGs is a clean problem, which can be 
calculated using CDM particles and gravity.  The genus topology is an important 
check that the N-body simulations here are doing a good job identifying 
the LRG galaxies.  


\begin{figure}
\center
\includegraphics[scale=0.45]{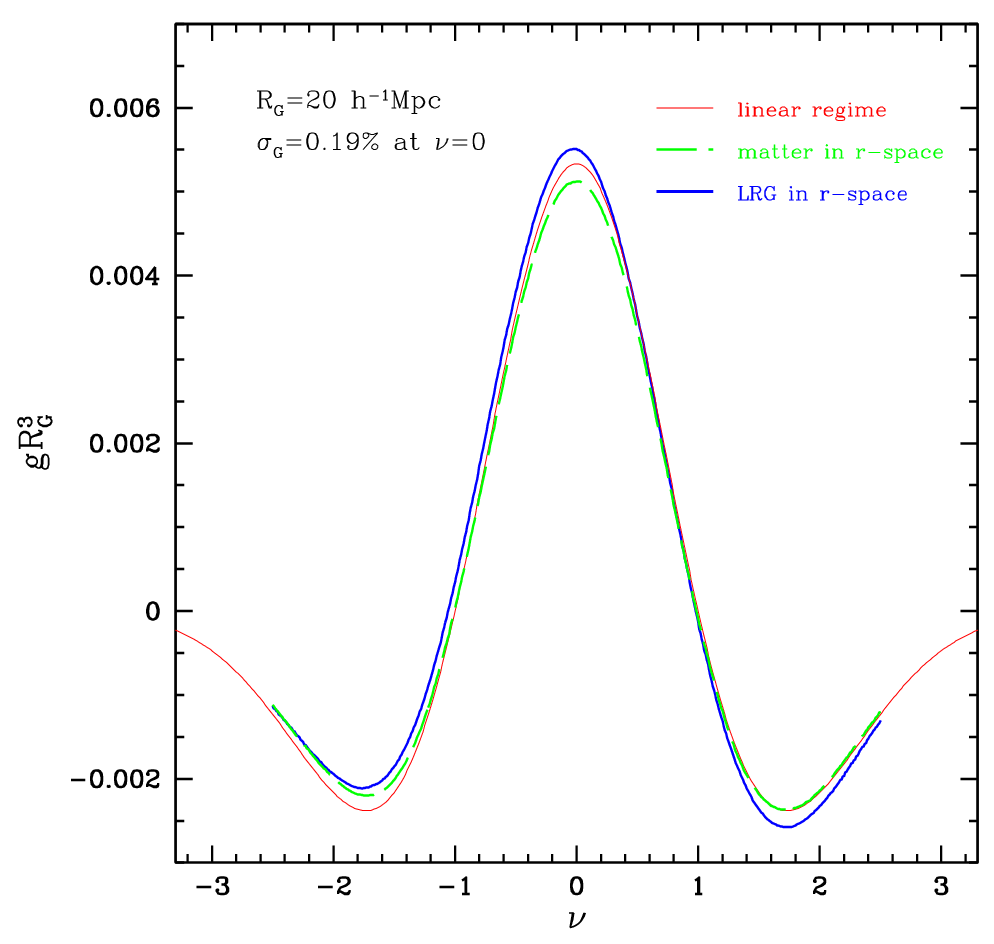}
\caption{ 
Genus curves of the matter density field (dashed line) and
mock LRG distribution (thick solid line) in real space at redshift $z=0$ calculated from
the whole simulation cube. The random phase Gaussian curve (thin solid line)
corresponding to the $\Lambda$CDM model we adopted
is plotted for a comparison.
The uncertainty in the genus at $\nu=0$ is 0.19\% for the mock LRG 
distribution.
}\label{fig-comp}
\end{figure}

\begin{figure}
\center
\includegraphics[scale=0.45]{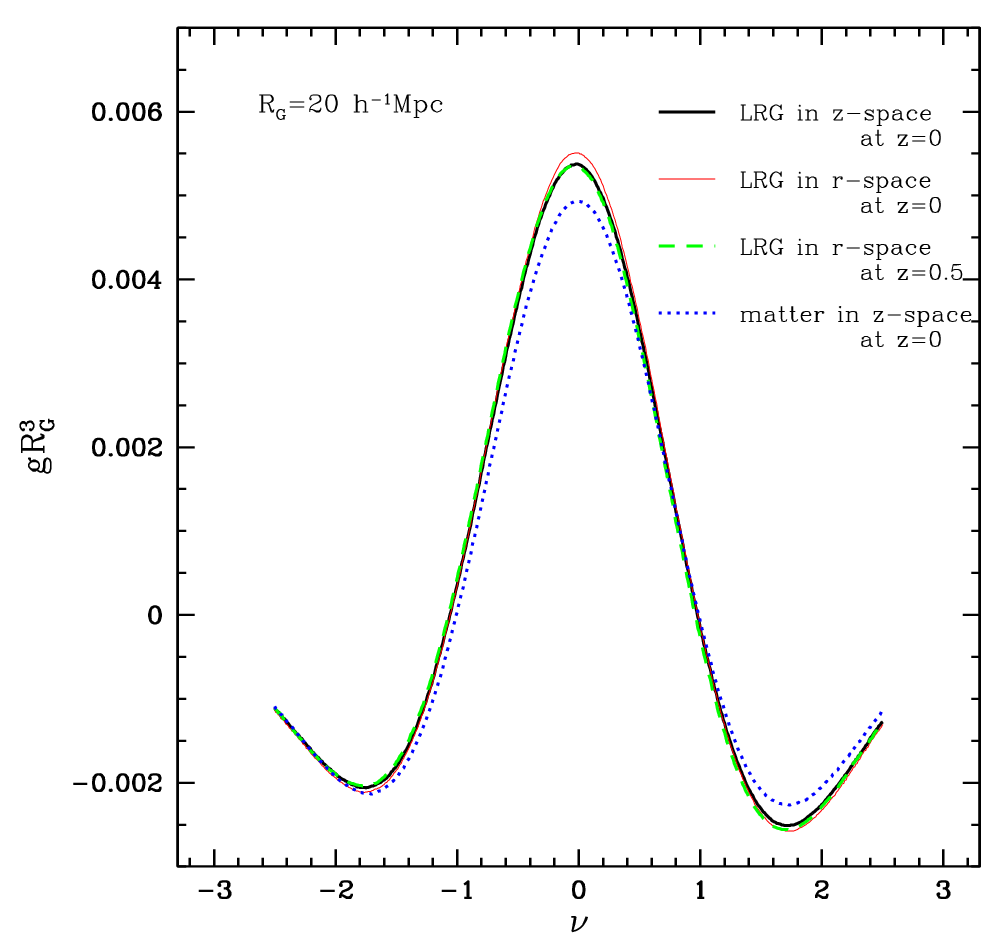}
\caption{ 
Genus curves of the mock LRG galaxies for the entire simulation cube 
in real space (thin solid line) and in redshift space (thick solid line) 
at redshift $z=0$.
The dotted line is the real sapce
genus curve of the matter density field in redshift space.
The genus curve of the mock LRGs at $z=0.5$ is also plotted with 
dashed line.
}\label{fig-comp}
\end{figure}

We now present the genus topology analysis of the Horizon Run.
Figure 13 compares the genus curves of the matter density field (dashed line) and
mock LRG distribution (thick solid line) calculated from the whole simulation cube using 
the periodic boundary condition.
The distribution of our mock LRG galaxies in the simulation 
cube is smoothed with a smoothing length of $20 h^{-1}$Mpc.  
The curves are compared with the random phase Gaussian curve
(thin solid line from $\nu=3.3$ to 3.3) corresponding
to the $\Lambda$CDM model we adopted (see Eq. 9).
Note that the LRG genus curve shows more deviations in the high-density regions
($\nu >1$) from the Gaussian curve
than the matter genus curve. This is because the mock LRGs are biased
objects, so we see more of them in the crowded regions relative to the background matter.
The overall amplitude of the LRG genus curve is very close to that of the linear one.

In Figure 14 we show the mock LRG genus curves for the entire cube in real 
(thin sold line) and redshift space (thick solid line) at redshift $z=0$.
The redshift space distortion is mimicked 
by adding the $x$-component of peculiar velocities of mock LRGs 
to their positions. The smoothing scale is again $20 h^{-1}$Mpc.
The dotted line is the matter genus curve in redshift space. 
It can be seen that the redshift space distortion effects make
the amplitude of the genus curve decrease. But the genus curve of
mock LRGs is much less affected by the redshift distortion 
when compared with that of matter field. 
This is again due to the strong statistical biasing of the LRG galaxies.
These curves are compared with the genus curve
of mock LRG galaxies in real space at redshift $z=0.5$ (dashed line).

\begin{figure}
\center
\includegraphics[scale=0.45]{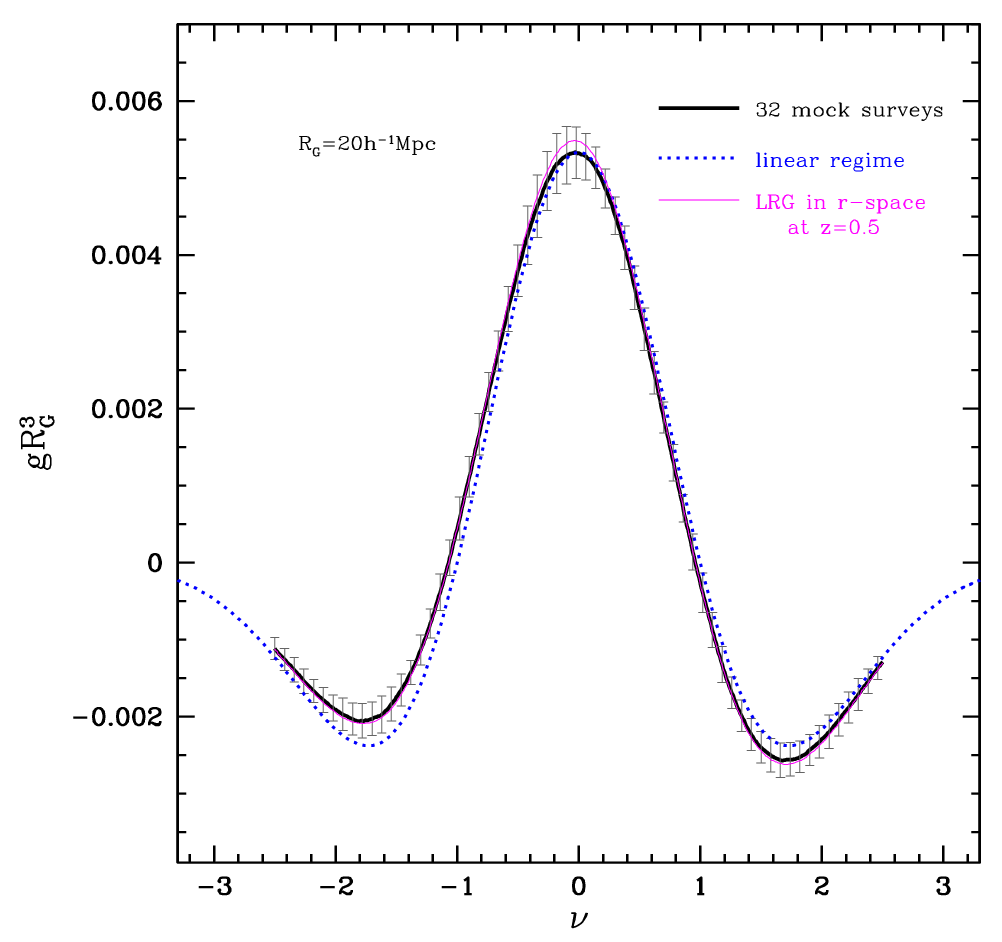}
\caption{
Mean genus curves of the mock LRG galaxies averaged over 32 mock surveys 
(thick solid line) with $3\sigma$ error bars.
The dotted line is the
genus curve of the linear regime of the matter field.
The genus curve of the mock LRGs at $z=0.5$ is also plotted with thin
solid line.
}
\label{fig-11}
\end{figure}

Figure 15 shows the mean genus curves averaged over 32 mock SDSS-III
LRG surveys. Each survey is a cone-shaped volume extending out to redshift 
$z=0.6$ along the past light cone of our Horizon Run,
and the whole survey cone is used to calculated the genus curve.
Peculiar velocities are added to radial distances
of LRGs to mimic the redshift distortion effects.
Since the mock LRGs have the mean density of 
$3\times 10^{-4} (h/{\rm Mpc})^3$ or a mean separation of 15$h^{-1}$Mpc, 
the smoothing length of
$20 h^{-1}$Mpc is large enough to beat down the shot noise effects.
Also shown are the real space genus curves of the mock LRGs  
at redshift $z=0.5$ and of the random phase Gaussian density field
corresponding to the $\Lambda$CDM model adopted in our
simulation. Note that there is no fitting involved.
The $3\sigma$ error bars are obtained from the genus distribution of 32 mock LRG surveys.

The genus curves of mock LRGs deviate from the linear regime curve
because of non-linear biasing and gravitational effects. 
Table 2 lists the genus-related parameters at the Gaussian smoothing scale of
$20h^{-1}$Mpc for LRGs measured from the whole
simulation cube and from the 32 mock SDSS-III surveys.
For the whole simulation cube results (the second and third columns)
the error bars are estimated based on the variance 
in these parameters seen in 64 quarter-sized subsamples.  
For mock SDSS-III LRG survey results
the estimates are the mean values averaged over 32 mock SDSS-III
surveys, and the error bars are for a single survey.
The linear regime predictions are also given for comparison.

%
%
%
%
%

\begin{deluxetable*}{c|cccc}
\tablecaption{Genus-related Parameters of the mock Luminous Red Galaxies
at $R_G = 20h^{-1}$Mpc}
\tablewidth{0pt}
\tablehead{
\colhead{Parameters}
&\colhead{Real Space}
&\colhead{Redshift Space}
&\colhead{Mock SDSS-III}
&\colhead{Linear Regime Prediction}
}
\startdata
$gR_G^3$ &$(5.496\pm0.007)\times10^{-3}$ &$(5.361\pm0.007)\times10^{-3}$ &$(5.362\pm0.080)\times10^{-3}$ & $5.33\times10^{-3}$ \\
$\Delta\nu$ &$-0.0427\pm0.0023$ &$-0.0451\pm0.0020$ & $-0.0427\pm0.0064 $& 0\\
$A_V$ &$0.830\pm0.002$ & $0.828\pm0.002$&$0.832 \pm 0.017 $ & 1\\
$A_C$ &$1.082\pm0.002$ & $1.083\pm0.002$&$1.088\pm0.023 $ & 1 
\enddata
\label{gentable}
\tablecomments{
Cols. (1) Genus parameter
(2) Mock LRGs in real space at $z=0.5$
(3) Mock LRGs in redshift space at $z=0.5$
(4) Mock LRGs in past light-cone space
(5) Linear regime predictions
}
\end{deluxetable*}

It can be seen that the amplitude of the observed genus curve of the mock
SDSS-III LRGs is only 0.6\% different from that of the linear theory expectation,
and will be determined with 1.5\% uncertainty. The non-Gaussian 
parameters $\Delta\nu, A_C,$ and $A_V$ show statistically significant deviations 
from the Gaussian curve, which is characteristic of the LRGs identified 
in our simulation.

\section{Conclusions}

Baryon oscillations are believed to be the method of characterizing 
dark energy ``least affected by systematic uncertainties'' according to the 
Dark Energy Task Force Report.   But the baryon oscillation is 
affected by all kinds of non-linear effects, so comparison 
with large N-body simulations is essential to control the systematics. 
Gott et al. (2009)'s study of the topology of the LRG galaxy clustering 
in the current SDSS shows that large N-body simulations 
are accurately modeling these non-linear gravitational and biased 
galaxy formation effects.

To aid in calibrating non-linear gravitational and biasing effects 
in the future Sloan-III baryon oscillation scale survey (BOSS), 
we have made here a large N-body simulation: $4120^3 = 69.9$ billion 
particles in a $(6592 h^{-1}$Mpc$)^3$ 
box, which models the formation of LRGs in this region.  
Using improved values over those adopted by the Millennium Run, which 
was based on WMAP-1 initial conditions (Spergel et al. 2003), our simulation 
has more power at large scales ($n_s = 0.953$ versus $n_s = 1$, which should 
make for a higher frequency of large scale structures), and
more accurate normalization at $8 h^{-1}$Mpc scale ($\sigma_8 = 0.796$
while Millennium Run used $\sigma_8=0.9$).

We have made 
32 mock surveys of the Sloan III survey, which should allow us to correct 
for small systematic effects in the measurement of the physical scale 
of the baryon oscillation peak as a function of redshift.  
We predict from our mock surveys that the BAO peak scale can be measured with 
the cosmic variance-dominated uncertainty of about 5\% when the SDSS-III sample
is divided into three equal volume shells, or about 2.6\% when a thicker shell
with $0.4<z<0.6$ is used.  We find that one needs to correct the scale for the
systematic effects amounting up to 5.2\% to use it to constrain the 
linear theories.

The amplitude of the genus curve measured in redshift space in the mock SDSS III
survey is, in the mean, expected to be equal to the linear regime prediction 
to an accuracy of about 2\% at the smoothing scale of 20 $h^{-1}$Mpc.
So the systematic evolution correction in the amplitude of the genus 
curve will be tiny.
Also, the RMS uncertainty in the amplitude in the genus curve due to the
cosmic variace as measured in one SDSS-III survey is only 1.5\% 
at 20 $h^{-1}$Mpc scale. Since the uncertainty scales with inverse 
square root of the number of resolution elements (or equivalently with
$R_G^1.5$), this implies the uncertainty of about 1.0\% at 15 $h^{-1}$Mpc scale,
which will be the mean separation of the SDSS-III LRGs.
In subsequent papers we shall show how 
measuring the amplitude of the genus curve at different smoothing length
makes possible an independent measure of scale complimenting the baryonic
oscillation bump method for measuring $w(z)$.

As dark energy makes up three quarters of 
the universe today, making such improvements in characterizing dark 
energy are particularly exciting.  
Much ancillary science can be done with this large N-body simulation; 
detailed modeling of the integrated Sachs-Wolf effect and comparison 
with counts of LRG galaxies in the sky, for example.  
We can search for Great Walls 
and voids to estimate their frequency and size (Gott et al. 2005).  
Here, our large volume size and WMAP-5 based 
initial conditions (Komatsu et al. 2008) should be of great benefit.  
We will make the simulation data available to the community
(See http://astro.kias.re.kr/Horizon-Run/ for information).  
Please cite this paper when the simulation data is used.

\acknowledgments
The Horizon Run Simulation was made on the TACHYON, a SUN Blade system, of 
Korea Institute of Science and Technology Information (KISTI)
which provided a full support of the computing resources.
JHK is supported by the Korean Research Foundation Grant KRF-2008-357-C00050 funded
by Korean government.
JHK thanks U.-L. Pen for his hospitality and the support of CITA.
CBP acknowledges the support of the Korea Science and Engineering
Foundation (KOSEF) through the Astrophysical Research Center for the
Structure and Evolution of the Cosmos (ARCSEC). JRG is supported by NSF grant
AST-0406713.  JD acknowledges NSERC for research funding.
We thank Korea Institute for Advanced Study for providing computing resources 
(KIAS linux cluster system) for this work.



\appendix
\section{A. The Starting Redshift and the number of time steps}
\label{sec:appendixA}
\begin{figure}
\center
\includegraphics[scale=0.45]{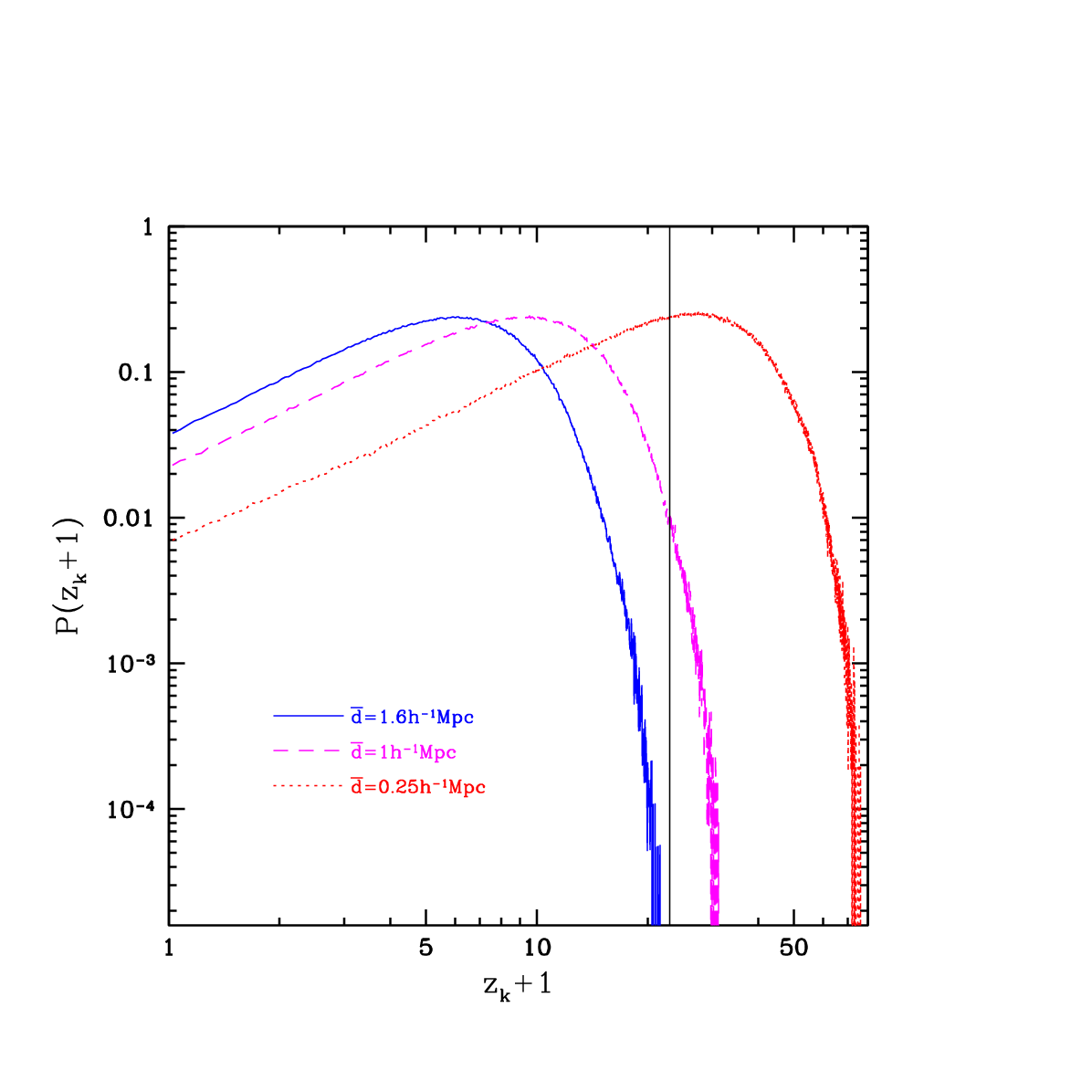}
\caption{
Distribution of the Zel'dovich redshifts (see text for its definition) 
for three different cases of pixel size.
The vertical bar marks the initial starting redshift adopted in our simulation.
}
\label{fig-a1}
\end{figure}

The TreePM code (Dubinski et al. 2004) used in this paper
adopts the first-order Lagrangian perturbation scheme
to generate the initial Zel'dovich displacement
(for comparison with the second-order scheme, refer to
Crocce et al. 2006 and references therein).
The starting redshift of the simulation should be chosen so that
the Lagrangian shifts of particles are not larger than
the pixel spacing.
This is because, if the initial displacement is larger than a pixel spacing,
a particle may not fully experience the local differential gravitational potential
(or Zel'dovich force)
and, consequently, the shift calculated by using the first-order scheme
could not fully reflect the fluctuations at the pixel scale.
To quantitatively estimate the proper starting redshift, we introduce
the Zel'dovich redshift, $z_k$, of a particle defined as the epoch when its
Zel'dovich displacement becomes equal to the pixel size either in $x$, $y$, or $z$ direction.
In Figure \ref{fig-a1},
we show the distribution function of the Zel'dovich
redshift, $z_k$, for three simulations tagged by the mean particle 
separation (${\bar d}$), or equivalently the pixel size.
To measure the distribution,
we perform an initial setting moving $256^3$ particles from the initial
conditions defined on a $256^3$-size mesh.
One should acknowledge a lack of the large-scale power
causing a possible underestimation of $z_k$ especially for the
small ${\bar d}$ simulation.
Each distribution shows a power-law increase with $z_k$
and a sharp drop after a peak.
It can be seen that no particle experiences a shift larger than a pixel size at $z=23$
when ${\bar d}=1.6h^{-1}$Mpc, which justifies our choice of the initial redshift.

\begin{figure}
\center
\includegraphics[scale=0.45]{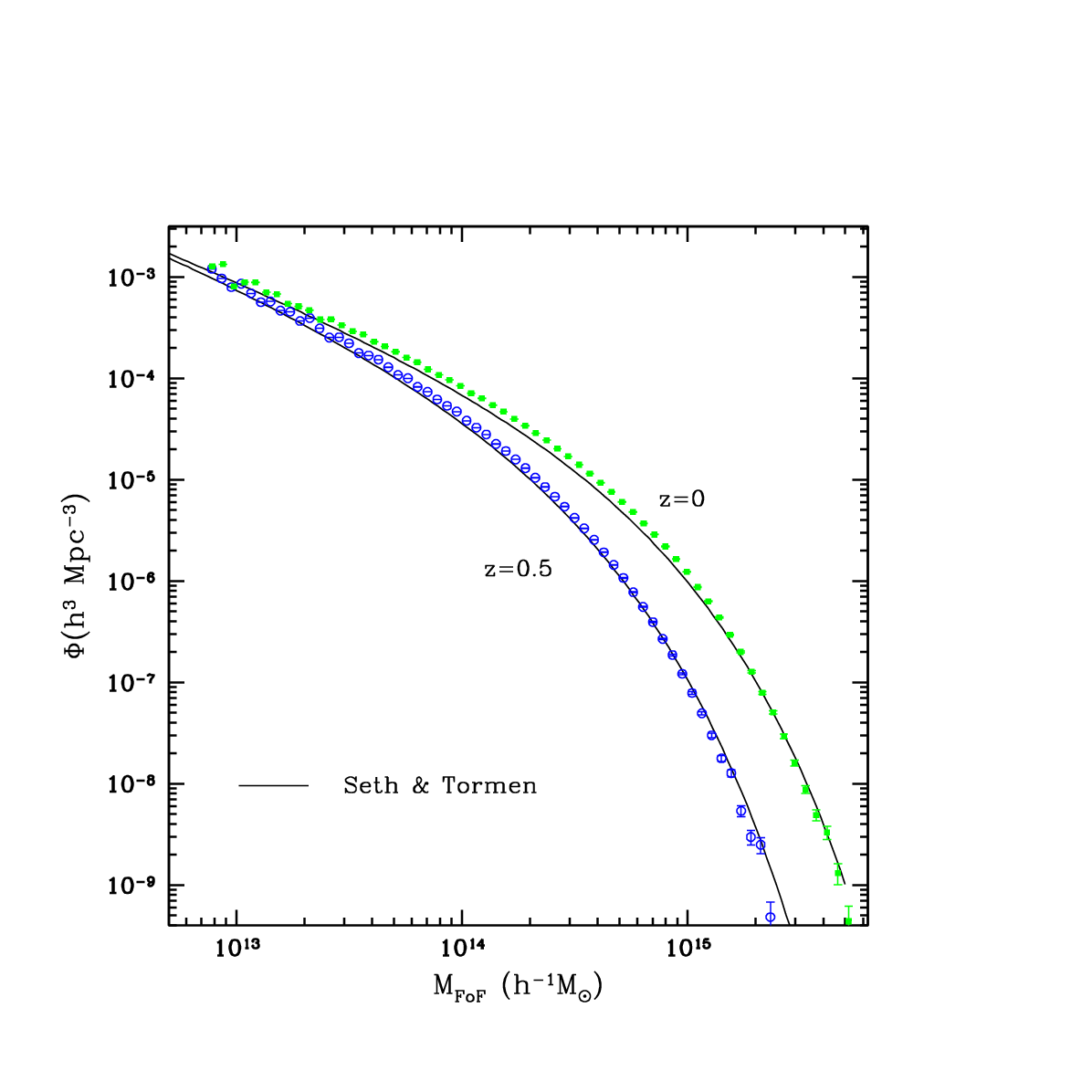}
\caption{
FoF halo mass functions. The filled squares and open cirlces
are the mass functions of the simulated FoF halos at z=0 and 0.5, respectively.
The solid curves show the Sheth \& Tormen (1999)'s
function.
}
\label{fig-a2}
\end{figure}
We adopted 400 time steps to evolve the Cold Dark Matter particles from $z=23$ to 0.
We set the force resolution to 0.1 $d_{\rm mean}$.
Here we present a test justifying our choice of the number of time steps.

Due to this relatively larger step size and lower force resolution,
one may raise a question about the simulation accuracy whether it may
have sufficient time and force resolutions to correctly model 
the formation of halos and subhalos.

One of the simplest ways to judge whether a simulation retains a sufficient power
to resolve small structures is to compare
the simulated FoF halo mass function with the well-known functions obtained
from very high resolution simulations
with a special emphasis on the population of less massive halos.
This is because coarse time-step size or poor force resolution may destroy
smaller structures more easily.
In Figure A2, 
Sheth \& Tormen (1999; {\it dotted})'s fitting function is 
shown at $z=0$ ({\it upper}) and 0.5 ({\it lower}).
The corresponding FoF halo mass functions from our simulation are shown by 
filled squares ($z=0$) and open circles ($z=0.5$).
This comparison demonstrates that the Horizon Run resolves structures with
mass down to about ($M_{30} \simeq 8.87\times10^{12} ~h^{-1}{\rm M_\odot}$,
which is the mass of 30 simulation particles. This mass limit is below 
the expected lower mass limit of the BOSS LRGs, and therefore formation of
the dark halos associated with the BOSS LRGs are accurately simulated by
the Horizon Run.

\section{B. Correlation Functions of Mock and SDSS LRG Galaxies}
\label{sec:appendixB}
To compare the CFs of the SDSS LRG and simulated mock LRG samples 
we built 24 mock SDSS LRG samples from eight all-sky 
past light cone data in a way that they do not overlap with one another.
We apply the same mask of the SDSS survey (i.e. the SDSS-I plus SDSS-II).
We use 
the SDSS LRG sample (S1; $-23.2 \le {\mathcal M}_{0.3g} \le -21.2$, $0.2 \le z\le0.36$) 
which is described in detail by Gott et al. (2009).
Here, ${\mathcal M}_{0.3g}$ is the absolute magnitude measured at $z=0.3$
(We omit the subscript 0.3 in Figure B1).
We re-calculate the comoving distance based on the WMAP 5-year cosmology
and obtain the mean number density of the LRGs 
as ${\bar n_{\rm LRG}}=9.1\times10^{-5} (h^{-1}{\rm Mpc})^{-3}$
which is slightly lower than the previous measurement,
${\bar n^\prime_{\rm LRG}} = 9.4\times10^{-5} (h^{-1}{\rm Mpc})^{-3}$
obtained by using the WMAP 3-year parameters.
As the number density of LRGs in the sample S1 is nearly constant over redshift 
(Zehavi et al. 2005), the lower-mass limit of the corresponding mock LRGs 
(dark halos) should be a function of redshift.
We apply the redshift-invariant multiplicity function of subhalos (or LRGs)
to find the relation between the critical mass and redshift.
This method allows a small fluctuation across 24 surveys
in the mean number density of mock galaxies
(${\bar n_{\rm mock}} = 9.1 \pm 0.2 \times10^{-5} (h^{-1}{\rm Mpc})^{-3}$).
Also we apply the redshift distortions to the mock galaxies to mimick observations.

\begin{figure}
\plotone{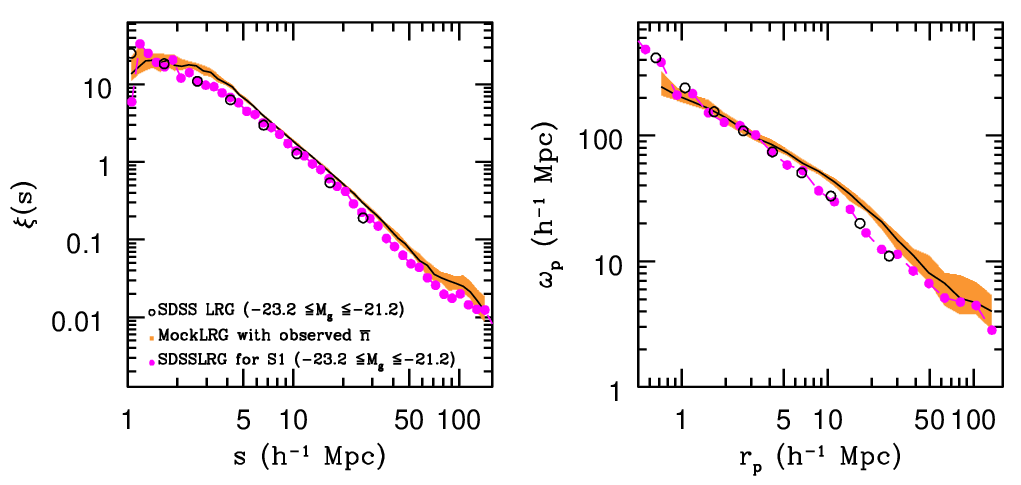}
\caption{
({\it Left}):
The three-dimensional CFs of mock LRG and SDSS LRG. The shaded region shows the
1--$\sigma$ distribution of CFs of the 24 mock surveys while the filled circles
are the measured CFs of the SDSS-II LRG sample.
The solid line marks the median value of the mock surveys.
For comparisons, we overplot those CFs of SDSS LRG subsamples under various magnitude limits
(Zehavi et al. 2005).
({\it Right}): Same as the left panel but for the projected CF.
}
\label{fig:fb1}
\end{figure}

The left panel of Figure \ref{fig:fb1} shows the average
CF of mock LRGs (solid line) together with the 1--$\sigma$ limits (shaded region)
measured on the past light cone surface.
Open circles are the CF of the SDSS LRGs calculated by Zehavi et al. (2005) 
based on the spatially-flat $\Omega_m=0.3$ cosmology 
and filled circles are the CF measured by us using the SDSS LRG data (S1)
based on the WMAP 5-year cosmology.
The error bars of the observational data should be similar to those of
the mock survey data.
The simulated LRG sample has a CF that impressively mimics the observations both in shape
and amplitude. But its amplitude seems slightly higher than 
that of the observation by 5\% -- 15\% on the intermediate scales 
($3 \la  s \la 50 h^{-1}{\rm Mpc}$).

The projected CF, $\omega_p(r_p)$, is another measure 
of the consistency between simulation and observation.
The projected distance ($r_p$) and the radial distance ($\pi$) between two LRGs
is related to each other by $r_p^2 + \pi^2 = s^2$ where $s$ is the three-dimensional distance 
between the pair in redshift space.
The projected correlation is obtained by integrating 
the two-dimensional CF $\xi(r_p,\pi)$ along the line of sight:
\begin{equation}
\omega_p(r_p) = \int_{-\pi_{\rm max}}^{\pi_{\rm max}} d\pi \xi(r_p,\pi),
\end{equation}
where $\pi_{\rm max}$ corresponds to the survey depth in the radial direction.
In the right panel of Figure \ref{fig:fb1}, we show the mean (solid line) and 
the 1--$\sigma$ limit (shaded region) of the projected CFs measured from the 
24 simulated LRG samples. 
These simulated CFs match the observed counterparts well 
except that it is slightly higher at $5 \la r_p \la 20 h^{-1}{\rm Mpc}$, 
which was also seen in the three-dimensional CF.
%
%
This deserves further analysis and will be investigated in a separate paper.

{}
\end{document}